\documentclass[12pt,english,a4paper]{article}

\pdfoutput=1

\usepackage{amsmath,amsfonts,amssymb,babel,slashed,color,empheq,tensor}
\usepackage{skak}
\usepackage{jheparxiv}
\usepackage{physics}
\usepackage[utf8]{inputenc}
\usepackage{lmodern}
\usepackage{latexsym}
\usepackage{mathrsfs}
\usepackage{braket}		
\usepackage{setspace}
\usepackage{stackengine} 
\usepackage{graphicx}
\usepackage[autostyle=true]{csquotes}
\usepackage[T1]{fontenc}
\usepackage{xcolor}
\usepackage[all]{xy}
\usepackage{tikz-cd}
\usepackage{float}
\usepackage{roundrule}   
\usepackage{mathtools}
\usepackage{twistor}
\usepackage{wasysym}

\usepackage{scalerel}
\def\rlwd{.9pt}
\def\lhexbrace{\kern1pt%
\setstackgap{S}{0pt}\def\stackalignment{l}
\ThisStyle{\scalerel*{%
  \stackunder[-\rlwd]{%
    \stackon[-\rlwd]{\roundrule{\rlwd}{4pt}}{\rotatebox{60}{\roundrule{4pt}{\rlwd}}}%
  }{\rotatebox{-60}{\roundrule{4pt}{\rlwd}}}%
}{\SavedStyle[}}}
\def\rhexbrace{%
\setstackgap{S}{0pt}\def\stackalignment{r}
\ThisStyle{\scalerel*{%
  \stackunder[-\rlwd]{%
    \stackon[-\rlwd]{\roundrule{\rlwd}{4pt}}{\rotatebox{-60}{\roundrule{4pt}{\rlwd}}}%
  }{\rotatebox{60}{\roundrule{4pt}{\rlwd}}}%
}{\SavedStyle[}}\kern1pt}

\makeatletter

\newcommand{\ib}{\boldsymbol{\mathtt{I}}}
\newcommand{\jb}{\boldsymbol{\mathtt{J}}}
\newcommand{\kb}{\boldsymbol{\mathtt{K}}}

\newcommand{\mso}{\mathfrak{so}}

\newcommand{\cT}{\mathcal{T}}

\newcommand{\diff}{\textnormal{d}}

\newcommand{\nn}{\nonumber}

 \def\one{\mbox{1 \kern-.59em {\rm l}}}

\newcommand{\cK}{\mathcal{K}}

\newcommand{\msf}[1]{\mathsf{#1}}

\def\a{\alpha}  \def\b{\beta}

 \def\L{\Lambda}

\makeatother

\begin{document}

 
\title{\Large Dynamical Covariant Quantum Spacetime \\ with Fuzzy Extra Dimensions in the IKKT model}

\author[1]{Alessandro Manta,}
\emailAdd{alessandro.manta@univie.ac.at}
\author[1]{Harold C. Steinacker,}
\emailAdd{harold.steinacker@univie.ac.at}
\affiliation[1]{Department of Physics, University of Vienna, \\
Boltzmanngasse 5, A-1090 Vienna, Austria}

\abstract{We consider general $k=-1$ FLRW covariant quantum spacetimes $\cM^{3,1} \times \cK$ with fuzzy extra dimensions $\cK$ as classical solutions of the IKKT matrix model. The coupled equations of motion are recast in terms of conservation laws, which allow to determine the evolution of spacetime in a transparent way.
We show that $\cK$ is stabilized as a classical solution in the presence of a large $R$ charge, corresponding to internal angular momentum.
This provides a  mechanism to maintain a large hierarchy between UV and IR scales. 
We also argue that
the evolution of spacetime is determined by a balance between classical and quantum effects, leading to a cosmic scale factor $a(t) \sim t$ and
constant dilaton  at late times. 
On such a background, the undeformed IKKT model leads to a higher-spin gauge theory including gravity.

}

\maketitle
\section{Introduction}

It is expected on general grounds that spacetime acquires some quantum structure at very short distances. 
Understanding the nature of quantum spacetime requires guidance from theory. We address this issue based on the IKKT matrix model \cite{Ishibashi:1996xs}, which is a leading candidate for such a fundamental theory, closely related to type IIB superstring theory. 

Although this matrix model is simple to write down, it is not evident how spacetime and gravity emerge. In recent years, evidence has mounted in non-perturbative studies \cite{Anagnostopoulos:2022dak,Chou:2025moy,Chou:2024sgk,Asano:2024def} that the model exhibits non-trivial saddle points,
including in particular the emergence of 3+1 large dimensions. Such saddle points can be considered as vacua or backgrounds defining spacetime. Fluctuations around them provide the physical degrees of freedom, propagating on the $(3+1)$-dimensional background at weak coupling\footnote{It is important to note that the notion of "weak coupling" only makes sense on some given background.}.
This scenario has been explored in recent years by studying the physics on various candidates for such backgrounds \cite{Steinacker:2019fcb}.
This weak-coupling approach focuses on $(3+1)$-dimensional emergent geometry, and should not be confused with holographic interpretations of $(9+1)$-dimensional target space geometry, which was recently studied in the polarized IKKT model \cite{Hartnoll:2024csr,Komatsu:2024bop,Komatsu:2024ydh}.

Covariant cosmological quantum spacetimes \cite{Sperling:2018xrm} are prototypical examples of noncommutative $(3+1)$d spacetimes, with cosmological features and supporting interesting field theory structures. They constitute classical saddle points for mass deformations of the IKKT Matrix Model \cite{Ishibashi:1996xs}. 
The one-loop physics of this model on such cosmological backgrounds has been studied in \cite{Battista:2023glw, Steinacker:2024huv, Manta:2024vol}, and gravity was shown to emerge in the presence of fuzzy extra dimensions $\cK$ \cite{Steinacker:2016vgf}.
However, stabilizing $\cM \times \cK$ requires adding quadratic and/or cubic terms to the matrix model, which breaks the supersymmetry of the IKKT model\footnote{It is interesting that the polarized IKKT model \cite{Hartnoll:2024csr} does include quadratic and cubic terms in the action, however at first sight the signs are not in the required range. This will be studied in detail elsewhere.}. For another approach to emergent cosmological spacetimes from IKKT, see \cite{Brahma:2021tkh, Brahma:2022dsd}.

In this work, we propose generalizations of such backgrounds that solve the classical IKKT equations of motion without any deformation.
We also provide a mechanism for the stabilization of dynamical extra dimensions, while preserving the $SO(1,3)$ isometry group of spacetime.

More specifically, we consider time-dependent generalizations of the cosmological $k=-1$ background $T^\mu=\a(\tau) t^\mu$, with dynamical extra dimensions $T^{\ib}=f(\tau)\cK^{\ib}$.
Here $t^\mu$ are certain $\mso(4,2)$ generators defining undeformed quantum spacetime \cite{Sperling:2018xrm}, and 
$\cK^{\ib}$ are suitable noncommuting matrices defining fuzzy internal dimensions. 

This work refines previous works as follows:
\begin{enumerate}
    \item In \cite{Battista:2023glw} the time evolution  of $\cM^{3,1}$
    was studied for a similar ansatz as above,
    for fixed $\cK$ at one loop. It was found that the classical action dominates
    at late times, leading to $\a(\tau) \to \e^{-\tau}$. However this led to an inconsistent picture where the dilaton approaches zero at late times corresponding to strong YM coupling, 
    invalidating the weak coupling assumption. The present paper provides a consistent treatment, properly taking into account the time evolution of $\a, \rho $ and $f$, and identifying a window for weak coupling.

\item In \cite{Steinacker:2024huv}, the dynamics of $\cK$ and the non-abelian Yang-Mills sector was studied at one-loop, for fixed background $\cM$. 
Stabilization required subtle quantum effects describing the interaction of $\cK$ and $\cM$.
This was extended in \cite{Manta:2024vol} to include the gravity sector. 
\end{enumerate}

In this paper, we consider the combined system of $\cM$ and $\cK$ at the classical level, and establish a mechanism for stabilizing $\cK$ as a classical solution with rotating extra dimensions, corresponding to non-vanishing $R$ charge. Related proposals have been discussed in a different setting in \cite{Iso:2015mva,Steinacker:2014eua,Berenstein:2015pxa}.
This provides a mechanism to stabilize a large hierarchy between UV scale set by $\cK$, and the IR scale set by the curvature of FLRW spacetime. Moreover, it justifies the setup  in \cite{Steinacker:2021yxt} for the emergence of gravity on $(3+1)$-dimensional spacetime $\cM$.

Nevertheless, we also find that the classical treatment cannot give the full story, because it would still lead to $\a(\tau) \sim \e^{-\tau}$ at late times, which implies 
 a decreasing dilaton and hence increasing YM coupling. Therefore quantum effects must be included.
We provide several arguments 
which point to a consistent picture where the spacetime evolves as $\a \sim \e^{-\frac{3}{4}\tau}$, which leads to an asymptotically constant dilaton and cosmic expansion\footnote{That conclusion is largely insensitive to the matter distribution, because emergent gravity is screened in the extreme IR regime; this will be discussed in more detail elsewhere.} rate $a(t) \sim t$.
The present scenario therefore 
leads to a physically reasonable cosmic evolution
where the effective gauge couplings, the KK masses and the Newton constant are
approximately constant. The full consistency of such a background and further details of the resulting physics remain to be studied in more detail.

To make the paper more readable, a list of symbols and notations is provided in Appendix \ref{notation}.

\section{Covariant Quantum Spacetime}

We start by briefly recalling the definition of covariant quantum spacetime as a background in the IKKT matrix model.
The IKKT model is uniquely specified by maximal supersymmetry, and given by the $SO(1,9)$-invariant action \cite{Ishibashi:1996xs}
\begin{align}\label{eq:SO(1,9)action}
    S=\frac{1}{g^2}\Tr\Big([{\bf T}^{A},{\bf T}^{B}][{\bf T}_{A},{\bf T}_{B}]
    +\bar{\Psi}\Gamma^{A}[{\bf T}_{A},\Psi]\Big)\,,\qquad A,B=0,1,\ldots,9 \ .
\end{align}
for 9+1 hermitian matrices ${\bf T}^{A}$ and matrix-valued Majorana-Weyl spinors $\Psi$. Indices $A$ are contracted with the Minkowski metric $\eta=\operatorname{diag}(-1,1,1,\cdots,1)$.

We will consider non-trivial backgrounds or "vacua" of this model defined by some matrix configuration ${\bf T}^A$ specified below. More specifically we consider quantized symplectic backgrounds in the semi-classical regime, so that we can identify commutators of matrices with Poisson brackets of functions. Then the fluctuations ${\bf T}^A \to {\bf T}^A + \cA^A$ define a noncommutative gauge theory, where $\cA$  are recognized as gauge fields and scalar fields governed by an effective  metric describing a $(3+1)$-dimensional FLRW spacetime (and similarly for the fermions, which we will ignore here).
For a systematic introduction see e.g. \cite{Steinacker:2024unq,Steinacker:2019fcb}; some basic semi-classical identities for that background geometry are given in Appendix \ref{basic-identities}, which are used throughout the paper.

\paragraph{Undeformed $\cM^{3,1}\times \cK$ background.} 

We will consider matrix backgrounds describing
 a product geometry $\cM^{3,1} \times \cK \subset \R^{1,9}$, where 
$\cM^{3,1}$ describes physical spacetime, and
$\cK$ describes some compact extra dimensions, embedded in $(9+1)$-dimensional flat target space. Algebraically, such backgrounds are defined by 
 matrix configurations 
\begin{align}\label{eq:background-2}
{\bf T}^{A}=
\binom{T^{\mu}\otimes \one_\cK}{\one_{\cM}\otimes \cK^{\ib}}
   \,,\qquad  \mu= 0,1, 2, 3\,,\qquad {\ib}=4,\ldots,9\,
\end{align}
acting on $\cH = \cH_\cM \otimes \cH_\cK$.
Specifically, we will consider covariant quantum spacetime, where $T^{\mu}$ is defined
in terms of some unitary 
"doubleton" minimal representation\footnote{Here $n=0,1,2,...$ could be any positive integer; the minimal case $n=0$ was discussed in \cite{Manta:2025inq}.}  
$\cH_\cM \equiv \cH_n$ of $SO(2,4)$  \cite{Gunaydin:1998sw}
as
\begin{align}\label{eq:background}
T^\mu  = \frac{1}{R} \cM^{\mu 4} \sim t^{\mu}, \qquad \mu = 0,\ldots,3
\end{align}
cf. \cite{Sperling:2019xar}. These generators are covariant under $SO(3,1)$, i.e. they satisfy $\Lambda^\mu_{\ \nu} t^\nu = U^{-1} t^\mu U$ for some unitary $U \in U(\cH_\cM)$. 
Then the algebra of matrices on $\cH_\cM$ can be interpreted as quantized algebra of (higher-spin-valued)    spacetime:
\begin{align}
    \End(\cH_\cM) \sim \cC(\cM^{3,1})\otimes \hs \ .
\end{align}
We will work in the semi-classical regime indicated by $\sim$, where commutators can be replaced by Poisson brackets, using the basic identities in Appendix \ref{basic-identities}. 
The higher-spin sector $\hs \cong \cC(S^2)$ describing harmonics on $S^2$ will mostly be suppressed; for more details see \cite{Manta:2025inq}.

Similarly, the generators $\cK^{\ib}$, acting on $\cH_\cK$, 
describe (functions on) some quantized compact space $\cK$, the simplest example being a fuzzy sphere.\\

The effective spacetime geometry defined by this background turns out to be a specific $k=-1$ FLRW spacetime with a Big Bounce \cite{Sperling:2019xar}, which can be parametrized by spacetime coordinates
\begin{align}\label{embedding-3d-hyperboloid}
    \begin{pmatrix}
     x^0\\
     x^1\\
     x^2\\
     x^3
    \end{pmatrix}
    &=R\cosh(\tau)\begin{pmatrix}
     \cosh(\chi)\\
\sinh(\chi)\sin(\theta)\cos(\varphi)\\
\sinh(\chi)\sin(\theta)\sin(\varphi)\\
     \sinh(\chi)\cos(\theta)
    \end{pmatrix}\,  \nn\\[1ex]
    x^4 &= R \sinh(\tau)   \ .
\end{align}
Here $\tau$ can be recognized as a time-like parameter
of the resulting FLRW geometry on spacetime $\cM^{3,1}$ \cite{Steinacker:2017bhb,Sperling:2019xar}.
The internal space $\cK$ plays the role of compact extra dimensions in the spirit of Kaluza-Klein, with a finite number of modes.

This matrix background is a solution of 
the classical equations of motion  $\square T^\mu = \frac{3}{R^2} T^\mu$ of the IKKT matrix model with mass term,  
which sets the scale $R$. However, it is not a solution of the equations of motion of the massless model
\begin{align}
    \square {\bf T}^A = 0, \qquad \square = [{\bf T}^A,[{\bf T}_A,.]] \ \equiv \square_{1,9} = \square_{1,3}+\square_6 \ .
\end{align}
Similarly, static internal compact spaces $\cK$ are classical solutions of the matrix equations of motions only in the presence of cubic and/or quadratic terms in the matrix model, which are forbidden in the IKKT model since they would break maximal SUSY\footnote{The polarized IKKT model \cite{Bonelli:2002mb,Komatsu:2024bop,Hartnoll:2024csr} does have some quadratic and cubic extra terms, however the sign of the mass term appears to be the opposite of what we need. Nevertheless, this is a possible avenue for future work.}. 

To resolve these issues, we will consider a more generic deformation of this background which incorporates  $k=-1$ FLRW spacetime with generic time-dependent scale function $\alpha(\tau)$ as in \cite{Battista:2022hqn},
and similarly for $\cK$.
This will allow to find classical solutions of the undeformed IKKT model.


\section{Dynamical Covariant Quantum Spacetime}
\label{sec:dynamical-covar}

We consider a more general 10d background of the type
\begin{equation}
    \cM^{3,1}\times_{\tau} \cK
\end{equation}
defined by 
 matrix configurations \eqref{eq:background-2}
allowing for a cosmological time $\tau$ dependence 
\begin{subequations}
\label{time-dep-background-matrices}
\begin{align}
 T^\mu &= \alpha(\tau)t^\mu  \label{spacetime-deformed}\\
   T_{\ib}^+&=f(\tau) \cK_{\ib}^+
   \label{K-rot-a}\\
   T_{\ib}^-&=\bar{f}(\tau) \cK_{\ib}^-,\ \ {\ib}=2,3,4
   \label{K-rot-b}
\end{align}
\end{subequations}
where $\cK^{\ib}$ define a (static) compact quantum space, and 
\begin{equation}
    \label{complex-TI-def}T^{\pm}_{\ib}=T_{2\ib}+iT_{2\ib +1},\qquad \cK^{\pm}_{\ib}=\cK_{2\ib}+i\cK_{2\ib +1} \ .
\end{equation}
This will allow to solve the classical IKKT equations of motion for the cosmological spacetime $\cM^{3,1}$, without adding a mass term by hand. We will be looking for solutions where $\a(\tau)$ and $|f(\tau)|$ are slowly varying on cosmic time scales, while the phase of $f(\tau)$ will be allowed to rotate with a UV time scale.  \\

Consider first the undeformed cosmological quantum spacetime background $t^\mu$, $\mu=0,1,2,3$, solving the massive equations of motion
\begin{equation}\label{undeformed-eom}
    \square_tt^\mu=\frac{3}{R^2}t^\mu \ .
\end{equation}
We assume that the internal generators commute with spacetime $[\cK^{\ib},t^\mu] = 0$,  and satisfy
\begin{equation}\label{def-Lambda}
    \square_\cK \cK^{\ib}
    =\Lambda\cK^{\ib}\ , \qquad
    \square_\cK  =  [K^{\ib},[K_{\ib},.]
\end{equation}
for some dimensionless $\Lambda>0$, absorbing the dimensionality in $f(\tau)$.
For simplicity we also assume\footnote{Note that, in general, $\cK^{\ib} \cK_{\ib}$ is not proportional to the identity: we will take this simplifying assumption, which applies to many prototypical examples of fuzzy compact spaces, e.g. fuzzy spheres, fuzzy tori or minimal squashed $\C P^2_N$ \cite{Steinacker:2014lma}. If this condition does not hold, then spacetime $\cM^{3,1}$ would acquire some non-trivial dependence on the internal space corresponding to some warped geometry, cf. \cite{Nishimura:2013moa}. 
}
\begin{equation}
\label{C-condition}
    \cK^{\ib}\cK_{\ib}=C^2\one
\end{equation}
with $C^2>0$.
The eigenmodes of $\Box_\cK$ in $\End(\cH_\cK)$ are denoted with 
\begin{align}
     \Box_\cK \Upsilon_{\L} = \mu^2_\L\, \Upsilon_{\L} \  .
 \label{Box-6-K-EV}
\end{align}
Then 
$\cK$ leads to a finite tower of KK modes on spacetime
\begin{align}
 \Box_6 \Upsilon_{\L} =
    |f|^2 \Box_\cK \Upsilon_{\L} = m^2_\L\, \Upsilon_{\L} \ , \qquad 
 \label{Box-6-K-EV}
\end{align}
with 
\begin{align}
    m_{\L}^2 = m_\cK^2 \,\mu^2_{\L}\,. \qquad m_\cK^2 = |f|^2 \ .
\end{align}
Here $\mu^2_{\L}$ characterizes the spectral geometry of $\cK$ and $m^2_\cK = |f|^2$ sets the KK mass scale.
To get interesting $(3+1)$d physics, this should be a UV scale, while the cosmic curvature scale on the spacetime background is obtained from 
\begin{equation}\label{cosmo-IR-mass}
    \square_{1,3}\sim m_{\textnormal{cosm}}^2:=  \frac{\alpha^2}{R^2}
\end{equation}
which should be in the far IR. 
We will exhibit a dynamical mechanism to establish a large hierarchy between these scales.

We thus consider a deformed background $(T^\mu, T^{\ib})$
given by \eqref{time-dep-background-matrices},
for which the classical equations of motion reduce to 
two differential equations for the functions $\a(\tau)$ and $f(\tau)$, as well as a constraint \eqref{constraint-K} for the $\cK^{\ib}$. Due to the  $SO(1,3)$ covariance of the $t^\mu$, this ansatz is expected to solve also the full quantum dynamics. 

\paragraph{Remark.}

The last remark is important and deserves some discussion (cf. Chapter 7.6.2. in \cite{Steinacker:2024unq}).
It is easy to see (from group theory) that the most general $\tau$-dependent  $SO(1,3)$ vector operator acting on $\cH_\cM$ can be written as
\begin{equation}
\label{most-general-vector-31}
    \alpha(\tau)t^\mu+ \b(\tau)x^\mu \ .
\end{equation}
As pointed out in \cite{Battista:2023glw}, this configuration is  gauge equivalent\footnote{There is one gauge inequivalent configuration, the light-cone combination $Z^\mu \sim t^\mu\pm \frac{x^\mu}{rR}$, which is degenerate, $[Z^\mu,Z^\nu]=0$, $Z^\mu Z_\mu=0$. We will not consider this any further.} to $\alpha(\tau)t^\mu$. This means that we can choose the gauge $\beta(\tau)=0$ for the background; it also implies that the full quantum equations of motion for $T^\mu$ have the structure
\begin{align}
\label{full-eom-covar}
     \Box_{1,9} T^\mu = \frac{\del \Gamma[T]}{\del T_\mu} =: \cT^\mu 
\end{align}
where  $\Gamma[T]$ is the quantum contribution to the (matrix) effective action. Since $\cT^\mu$ enjoys the same $SO(1,3)$ covariance as $T^\mu$, it must have the form \eqref{most-general-vector-31}, just like $\Box_{1,9} T^\mu$. Therefore the full quantum dynamics can be solved in terms of two (modified) equations for the two functions $\a(\tau)$ and $ f(\tau)$, as well as a constraint to eliminate $\b$.

\subsection{Classical equations of motion}
Now consider the background \eqref{time-dep-background-matrices}.
We will denote $()' \equiv \frac{d}{d x_4}$, 
which amounts to a time derivative on the present backgrounds. Define 
(cf. (58) in \cite{Battista:2023glw})
\begin{align}\label{vareps-def}
\varepsilon:= \a^{-1}\frac{d\a}{d\tau} = x_4 \frac{\a'}{\a} \   \stackrel{!}{=} O(1)
\end{align}
which identifies the time scale 
of the cosmic background.
It is also convenient to define
\begin{align}
A = \a^2 \ .
\end{align}
The (bosonic) IKKT equations of motion are
\begin{equation}
    \square_{1,9}{\bf T}^A=0,\qquad \square_{1,9}=[{\bf T}^B,[{\bf T}_B,\ \cdot\ ]]=\square_{1,3}+\square_6 \ .
\end{equation}
These equations of motion are evaluated in the semi-classical regime in Appendix \ref{eom-computation}. 
The equation governing the evolution of $\cK$ takes the form
\begin{equation}\label{eomK}
    \boxed{(x_4^2+R^2)Af^{\prime\prime}+4Ax_4f^\prime+(x_4^2+R^2)A^\prime f^\prime+R^2\Lambda |f|^2f=0}
\end{equation}
while the equation governing the evolution of $\cM$ takes the form
\begin{align}\label{eomM}
    &\boxed{3A+3A^\prime x_4+\frac{1}{2}A^{\prime\prime}x_4^2+\frac{1}{2}R^2A^{\prime\prime}-C^2r^2R^2|f^\prime|^2=0 \ . }
\end{align}
There is also a constraint 
\begin{align}
\label{constraint-K}
     \boxed{[\cK^{\ib -},\cK_{\ib}^+](\bar{f}f^\prime-f\bar{f}^\prime)=0}
\end{align}
which arises from eliminating contributions proportional to $x^\nu$ in the eom (cf. \eqref{most-general-vector-31}); its role will become clear later.

We can rewrite these eom conveniently in terms of the cosmological time coordinate $\tau$, in the late $\tau$ regime
\begin{equation}
    x_4=R\sinh\tau\approx R\e^\tau, \ \ \partial_\tau\approx x_4\partial_{x_4}
\end{equation}
as
\begin{equation}\label{boxtmu}
    \boxed{0 = 3A+\frac{5}{2}\dot{A}+\frac{1}{2}\ddot{A}-r^2\e^{-2\tau}|\dot{f}|^2C^2}
\end{equation}
and
\begin{align}\label{boxk}
    &\boxed{0 = A\ddot{f}+3A\dot{f}+\dot{A}\dot{f}+R^2|f|^2f\Lambda}
    \ = (\partial_\tau + 3)(A \dot{f}) +R^2|f|^2f\Lambda \ .
\end{align}
Here and in the following we denote $\tau$ derivatives $\dot A \equiv \frac{d A}{d \tau}$ by dots.

\subsection{Solutions of the eom}

We now discuss solutions of these equations, starting with the solution for $\cM$, with fixed or vanishing $\cK$, and including the dynamics of $\cK$ in a second step.

\subsubsection{Purely $(3+1)$-dimensional solution}

Assume $f=0$ i.e. no $\cK$. Then the late time equation of motion for spacetime is
\begin{equation}
    3A+3A^\prime x_4+\frac{1}{2}A^{\prime\prime}x_4^2=0
\end{equation}
which is solved by 
\begin{equation}
    \alpha(x_4)=\frac{R\alpha_0}{x_4}\sqrt{1+\frac{R\alpha_1}{{x_4}}},\ \ \alpha_0,\alpha_1\in\mathbb{R} \ .
\end{equation}
For $\alpha_0\neq 0$, at late times this is dominated by $\alpha \sim e^{-\tau}$.
This behavior will be modified both by the 
dynamics of $\cK$ as well as quantum effects.
We will thus consider more general evolutions of the background, which will typically scale as
\begin{align}
    \a(\tau) \sim  e^{\varepsilon\tau}, \qquad -1 < \varepsilon < 0
\end{align}
at late times.

\paragraph*{Exact Solution.}
The exact semi-classical equation of motion for $f=const$, without late-time approximation, is solved by\footnote{Note that "in the very early universe" $x_4<R$, the equations of motion are solved by
\begin{equation}
  \alpha(x_4) =\frac{R\alpha_0}{x_4^2+R^2}\sqrt{R^2-x_4^2+R\alpha_1{x_4}},\ \ x_4<R \  \nonumber
\end{equation}
For generic $\alpha_0,\alpha_1$ the solution is not smooth at  $x_4=0$ and $x_4=R$.}
\begin{equation}
\label{alpha-for-static-f}
    \alpha(x_4) =\frac{R\alpha_0}{x_4^2+R^2}\sqrt{x_4^2-R^2+R\alpha_1{x_4}},\ \ x_4>R \ .
\end{equation}
Note that for "small" $x_4 = O(R)$, the 
semi-classical approximation is only justified for representations with very large $n \gg 1$.

\subsubsection{Dynamical $\cK$}

Since the internal Laplacian $\Box_6$ is a positive definite operator, any initially static $\cK$ is going to collapse rapidly, at least at the classical level. 
An intuitive mechanism to stabilize $\cK$ is to let it rotate in the transversal space $\R^6$; this amounts to giving it a non-vanishing $R$ charge as discussed in Section \ref{sec:R-current}, cf. \cite{Steinacker:2014eua,Iso:2015mva}.
In the present setting of cosmological spacetimes, this ansatz is perfectly reasonable, since the corresponding current is a homogeneous time-like vector field compatible with the isometries of the FLRW spacetime.
However $\cK$ has to satisfy some constraints, to avoid radiative effects.
We will therefore be working with the ansatz \eqref{K-rot-a}, \eqref{K-rot-b}
\begin{equation}\label{rotating-ansatz}
    T_{\ib}^+=f(\tau)(\cK_{2\ib}+ i\cK_{2\ib+1})=f(\tau) \cK_{\ib}^+,\quad T_{\ib}^-=\bar{f}(\tau)(\cK_{2\ib}- i\cK_{2\ib+1})=\bar{f}(\tau) \cK_{\ib}^-
  \end{equation}
for ${\ib}=2,3,4$, with fixed matrices $\cK_{\ib}$ and a complex time-dependent function
\begin{equation}
    f(\tau) = \chi(\tau)e^{i\omega(\tau)} \ .
\end{equation}
This describes $\cK$ rotating along an internal $U(1)\subset SO(6)$, which is the R-symmetry group of the model acting on the internal matrices 
\begin{equation}
    T_{2\ib}=\frac 12 (T^+_{\ib}+T^-_{\ib}),\ \qquad T_{2{\ib}+1}=\frac 1{2i} (T^+_{\ib}-T^-_{\ib}) \ .
\end{equation}  
 Consider first the constraint
\eqref{constraint-K}
\begin{equation}
    [\cK^{\ib +},\cK^-_{\ib}](f\dot{\bar{f}}-\bar{f}\dot{f})=0 \ ,
\end{equation}
which ensures that no Yang-Mills radiation $J_\mu\sim [K^{\ib},D_\mu K_{\ib}]$ is generated by the rotation of $\cK$, cf. \cite{Steinacker:2014eua}.
It can be solved in two ways:
\begin{itemize}
    \item Real\footnote{Up to an irrelevant constant phase.} $f$: this case is discussed in some detail in Appendix \ref{nonrotatingsol}. This  is physically not  satisfactory, since it leads to a rapidly oscillating radius of $\cK$, and hence oscillating physical moduli such as Kaluza-Klein (KK) masses. 
    
    \item $[\cK^{\ib +},\cK^-_{\ib}]=0$, which will arise again in the following. We will show that this leads to a non-trivial rotating solution $\dot{\omega}\neq 0$ which stabilizes $\cK$, and protects a large and stable
hierarchy between UV and IR scales.  We will focus on this case in the following.

\end{itemize}

\paragraph{Stabilization of the radius $\chi$.}

Using the time-derivatives of 
$f=\chi(\tau)\e^{i\omega(\tau)}$
\begin{align}
    \dot f&=(\dot\chi+i\chi\dot\omega)\e^{i\omega}\\
    \ddot f&=(\ddot\chi-\chi\dot\omega^2+2i\dot\chi\dot\omega+i\chi\ddot\omega)\e^{i\omega} \ ,
\end{align}
the equation of motion \eqref{boxk} for $\cK$ can be rewritten as the system of real equations
\begin{align}\label{chi-omega-eqns}
    &A\ddot\chi-A\chi\dot\omega^2+(3A+\dot A)\dot \chi+R^2\Lambda\chi^3=0  \\
    & \frac{d}{d\tau}(A \dot\omega) 
      = - (2\dot\chi \chi^{-1} + 3) A  \dot\omega \ .
\end{align}
The second equation  will be understood as 
$R$-current conservation law \eqref{j-explicit}. We can solve it exactly:
 \begin{align} 
   \frac{1}{A\dot\omega}\frac{d}{d\tau}(A \dot\omega) = \frac{d}{d\tau} \ln(A\dot\omega)
      &= - (2\dot\chi \chi^{-1} + 3) 
      = -(2\frac{d}{d\tau} \ln\chi + 3) \nn\\
 \frac{d}{d\tau} \ln(A \chi^2\dot\omega)  &= -3 
\end{align}
leading to $\ln(A \chi^2\dot\omega) = -3 \tau + \tilde{\ell}$,
 and hence
  \begin{align}  
  \label{omegadot-J-solution}
  \boxed{
  \dot\omega = \frac{\ell}{(\a\chi)^2} e^{-3 \tau} \ 
  }
\end{align}
for some constant $\ell$ of mass dimension $[{\rm mass}^2]$.

\paragraph{Large $\dot\omega$ regime.}

Consider the regime where $\dot\omega \gg  \frac{\dot{\chi}}{\chi}$ is very large, so that the rotational energy of $\cK$ dominates. Then we can simplify the first equation in \eqref{chi-omega-eqns} as
\begin{align}
    A\chi\dot\omega^2 = R^2\Lambda \chi^3 \ ,
\end{align}
which using \eqref{omegadot-J-solution} becomes 
\begin{align}\label{chi-A-relation}
\chi =  \left(\frac{\ell^2}{ R^2\Lambda}\right)^{\frac{1}{6}}  e^{-\tau} A^{-\frac{1}{6}} \ .
\end{align}
As expected,  this is slowly evolving at cosmic time scales. The $\cK$ mass scale is set by the prefactor, which is assumed to be large i.e. UV scale. 
Plugging this into the eom for spacetime $T^\mu = \a t^\mu$ \eqref{boxtmu} leads to 
\begin{align}
    &3A+\frac{5}{2}\dot{A}+\frac{1}{2}\ddot{A}
     = r^2C^2 (R^2 \Lambda)^{\frac{1}{3}} \ell^{-\frac{4}{3}} e^{-6\tau}  A^{-\frac{5}{3}}
\end{align}
This classical equation is expected to be modified at one-loop due to vacuum energy contributions, as discussed in Section \ref{one-loop}.

\paragraph*{Asymptotic solutions.} \label{asymptotic-sols}
Due to the $\e^{-2\tau}$ factor in \eqref{boxtmu}, we expect that the solution approaches the exponential time dependence  $\alpha\sim \e^{-\tau}$ of the 
background without $\cK$ for late times (recall $A = \a^2$), cf.
\eqref{alpha-for-static-f}, as will be confirmed numerically.
However, in order to understand the dynamics of $\cK$ more generally (and to take into account quantum corrections), we consider a more general ansatz for the evolution of the spacetime scale 
\begin{align}
    \alpha=\alpha_0 \e^{\varepsilon\tau}
\end{align}
with $\varepsilon = O(1)$, varying on a cosmic (IR) time scale. 
Then the equation of motion for $\chi$ is
\begin{equation}
    \ddot{\chi}-\chi\dot\omega^2+(2\varepsilon+3)\dot\chi=-\frac{R^2}{\alpha_0^2}\Lambda\e^{-2\varepsilon\tau}\chi^3 \ .
\end{equation}
Focusing on solutions for which 
$\chi$ varies only on cosmic time scales and the $\dot\omega^2$ term dominates on the lhs, this reduces to
\begin{equation}
    \dot\omega^2\approx \frac{R^2\Lambda}{\alpha_0^2}\e^{-2\varepsilon\tau}\chi^2
\end{equation}
both sides being UV scales.
Replacing $\dot\omega$ using $R$ charge conservation \eqref{omegadot-J-solution} 
leads to 
\begin{align}
    \chi&=\left(\frac{\ell^2}{\alpha_0^2\Lambda R^2}\right)^{\frac{1}{6}}\exp\left[-\left(1+\frac{\varepsilon}{3}\right)\tau\right] \ .
\end{align}
For $\varepsilon\neq -\frac{3}{4}$, solving equation \eqref{omegadot-J-solution} for $\dot\omega$ then gives
\begin{align}
    \omega&
    =\left(\frac{\ell^2R^4\Lambda^2}{\alpha_0^8}\right)^{\frac{1}{6}}\frac{1}{-1-\frac{4}{3}\varepsilon}\exp\left[-\left(1+\frac{4}{3}\varepsilon\right)\tau\right] \ .
\end{align}
We observe that $\dot \omega$ evolves on a cosmic time scale set by $\varepsilon$: it
is exponentially increasing for $\varepsilon < -\frac{3}{4}$, 
exponentially decreasing for $\varepsilon > -\frac{3}{4}$, and constant for $\varepsilon = -\frac{3}{4}$.
We will see that the dilaton (and hence the Yang-Mills coupling) behaves in the opposite way, which suggest that the critical case  $\varepsilon=-\frac{3}{4}$ is preferred, leading to a linear behavior for $\omega$:
\begin{align}
 \omega &=\left(\frac{\ell^2R^4\Lambda^2}{\alpha_0^8}\right)^{\frac{1}{6}}\tau =: \omega_0 \tau
     \qquad \qquad \mbox{for} \quad \varepsilon=-\frac{3}{4} \ , \nonumber\\
    \chi&=\left(\frac{\ell^2}{\alpha_0^2\Lambda R^2}\right)^{\frac{1}{6}}\exp\left(-\frac{3}{4}\tau\right)  
    \ .
\end{align}
Hence $\cK$ rotates in internal space with constant frequency $\omega_0$ (assumed to be a UV scale),
while the amplitude scales consistently with spacetime:
\begin{align}
\label{chi-alpha-34}
\boxed{
    \chi(\tau) \sim \a(\tau) \qquad \mbox{for} \quad \varepsilon=-\frac{3}{4} \ . 
    }
\end{align}
This result is interesting, because the combined $(9+1)$-dimensional matrix background \eqref{eq:background-2} then acquires a $\tau$-dependent overall factor ${\bf T}^A \to \a(\tau){\bf T}^A$. This  implies that the 1-loop effective action for the IKKT model coincides with that on the undeformed background $\a=1$,
since it is invariant under such a rescaling\footnote{Locally, for $\a\approx const$.}. 
Then there is no need to worry about complications from the extra $\hs$ components (as discussed in Section \ref{eff-frame-metric}) in the loop computation.
The case $\varepsilon=-\frac{3}{4}$ will turn out to be preferred for other reasons as well.

In any case, we obtain the KK mass scale  as
\begin{align}\label{KK-mass}
    m_\cK^2 = |\chi|^2 = \left(\frac{\ell^2}{\alpha_0^2\Lambda R^2}\right)^{\frac{1}{3}}\exp\left[-\left(2+\frac{2\varepsilon}{3}\right)\tau\right]
\end{align}
which is assumed to be a UV scale, slowly varying along the cosmic evolution $\a(\tau)$. Hence $m_\cK$ can be considered as constant on sub-cosmological scales.
This leads to 
a large separation of scales\footnote{The physical mass scales contain an extra factor  $\rho^{-2}$, cf.  \cite{Manta:2024vol} eq. (2.43), which drops out in the ratio.}  as long as
\begin{equation}
\label{hierarchy}
    \frac{m_{\cK}^2}{m_{\textnormal{cosm}}^2}=R^2\frac{m_{\cK}^2}{\alpha^2}=\left(\frac{\ell^2R^4}{\alpha_0^8\Lambda}\right)^{\frac{1}{3}}\exp\left[-\left(2+\frac{8\varepsilon}{3}\right)\tau\right]\gg 1
\end{equation}
where $m_{\textnormal{cosm}}$ is the cosmological IR mass scale \eqref{cosmo-IR-mass},
i.e. for
\begin{equation}\label{large-c-uv-ir}
    \frac{\ell^2R^4}{\alpha_0^8\Lambda}\gg 
    \exp\left[\left(6+{8\varepsilon}\right)\tau\right]
\end{equation}
hence for large internal "angular momentum" $\ell$. Note that this hierarchy is independent of time for $\varepsilon = -\frac{3}{4}$.

The same value for $\varepsilon$ is also selected on independent grounds.
We will see in Section \ref{dilaton} that the dilaton scales as $\rho^2 \sim e^{(3+4\varepsilon)\tau}$. Therefore an non-decreasing dilaton (ensuring weak coupling at late times) requires $\varepsilon\geq -\frac{3}{4}$.
On the other hand, maintaining a UV-IR separation at arbitrarily late times with this mechanism requires $\varepsilon\leq -\frac{3}{4}$. Combining these "constraints" uniquely suggests $\varepsilon=-\frac{3}{4}$. Although this is not a solution of all the classical equations of motion, we will argue that 
quantum effects should lead to this solution.


\paragraph{Classical vacuum evolution.}

\begin{figure}[h!]
    \centering
    \includegraphics[width=0.9\linewidth]{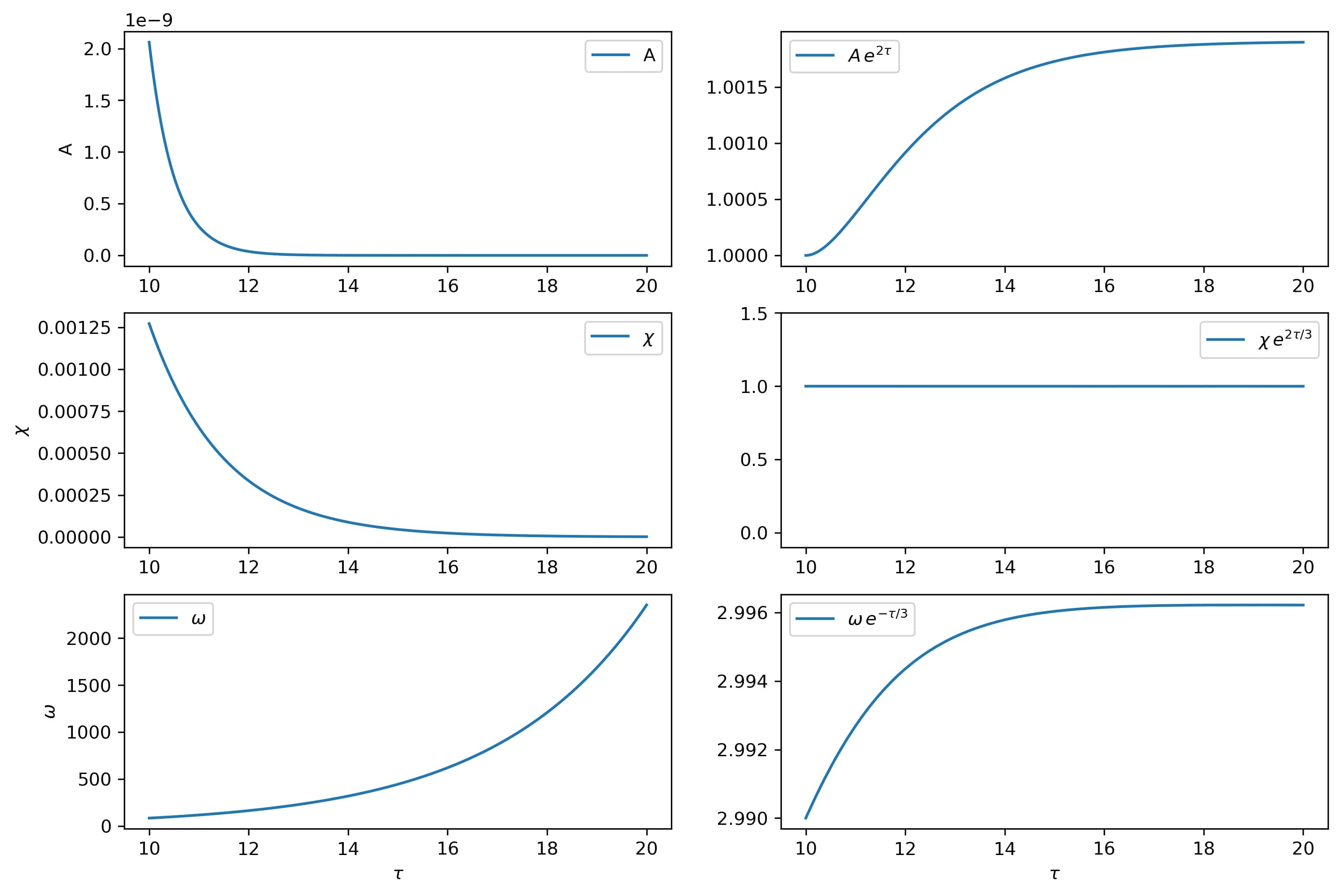}
    \caption{The behavior of $A(\tau)=\alpha^2(\tau)$, $\chi(\tau)$, $\omega(\tau)$, and their ratio with the asymptotic exponential scaling are shown.}
    \label{fig:osc-sol}
\end{figure}

Now we return to the classical spacetime solution, with asymptotic behavior $\varepsilon=-1$.
The stabilization mechanism for $\cK$ 
based on the non-vanishing $R$ charge applies, with a slow decay of the radius $\chi$ according to the cosmic expansion:
\begin{align}
     \chi&=\left(\frac{\ell^2}{\alpha_0^2\Lambda R^2}\right)^{\frac{1}{6}}\exp\left(-\frac{2}{3}\tau\right)\\
    \omega&=3\left(\frac{\ell^2R^4\Lambda^2}{\alpha_0^8}\right)^{\frac{1}{6}}\exp\left(\frac{\tau}{3}\right)
\end{align}
accompanied with an accelerated internal rotation at late times.
This demonstrates that there are classical solutions of the supersymmetric matrix model with non-trivial stable fuzzy extra dimensions $\cK$, which lead to a large hierarchy between UV and IR scales on an FLRW spacetime, with a reasonable local structure and long lifetime\footnote{The presence of exponentials and logs is an artifact of the coordinate choice $\tau$. The physical FLRW time $t$ is related to $\tau$ by \eqref{t-tau-rel-cosm}, so that all quantities scale with some power of $a(t)$.}. 
This behavior is confirmed by solving the classical equations of motion numerically, 
as shown in figure \ref{fig:osc-sol}.

 However, a background with $\varepsilon \approx -1$ is physically problematic, since the spacetime metric becomes degenerate for $\varepsilon = -1$, see Section \ref{eff-frame-metric}. Small deviations due to $\cK$ might help as discussed below, however the dilaton would approach zero at late times, which means that the gauge theory becomes strongly coupled, invalidating the weakly coupled approach.  

We can qualitatively assess subleading deviations from this
degenerate case for
classical solutions for $\alpha$, by making an ansatz $A(x_4)=x_4^{-2}h(x_4)$ with small $h$. Then the  equation of motion for $A$ gives
\begin{equation}
\label{h-eom-1}
    \frac{\diff(\dot{h}\e^\tau)}{\diff\tau}>0 \ .
\end{equation}
An exponential ansatz $h=\e^{\delta \tau}$ (locally) with $|\delta|\ll 1$ requires $\delta >0$,
i.e. it pushes the evolution to be "sub-critical", with
\begin{equation}
    \varepsilon>-1 \ .
\end{equation}
This is consistent with the numerical results in fig. \ref{fig:osc-sol}.
Thus the degenerate classical solution $\varepsilon = -1$ is avoided, leading to a well-defined $(3+1)$-dimensional spacetime.
Nevertheless,
the time dependence will be modified when quantum effects and matter are taken into account, as discussed in the following.

\subsection{Stabilization of spacetime and dilaton via quantum effects}\label{one-loop}

So far we have ignored quantum effects, which will modify the equation of motion for spacetime. As discussed around \eqref{full-eom-covar}, quantum effects are expected to preserve covariant quantum spacetime backgrounds with the structure $\a(\tau) t^\mu$, because these are the only $SO(1,3)$ vector operators (up to gauge invariance). While the purely classical analysis suggests $\varepsilon=-1$, we now provide qualitative arguments that quantum effects will modify this towards a physically preferred value of $\varepsilon=-\frac{3}{4}$;
the energy-momentum conservation law discussed in Section \ref{sec:e-m-tensor} 
should provide a suitable framework to confirm these effects.

Combining \eqref{boxtmu} with  \eqref{full-eom-covar}, the quantum equation of motion $\Box_{1,9} T^\mu = \cT^\mu$  
acquires an extra term on the rhs, with the structure
\begin{equation}
    \label{A-eom-mass-1-loop}
    \frac 1{2R^2}\big(
\ddot{A}+5\dot{A}+6A 
- 2C^2r^2\e^{-2\tau}|\dot{f}|^2\big)= 
     F(\rho) g^2 m_{\mathcal{K}}^2
     + O(g^4) \ 
\end{equation}
where  $A = \a^2$ is dimensionless, $g^2$ is the matrix model coupling constant, $F(\rho)$ some function of the dilaton $\rho$ due to geometric factors.
The lhs is obtained from $\Box_{1,9} T^\mu$, while the rhs is expected for $\cT^\mu$, assuming that the UV (Kaluza-Klein) scale $m_\cK$ governs the loop computation, taking into account \eqref{most-general-vector-31}.
 For a related computation see (6.28) in \cite{Steinacker:2024huv}. 
Using the result \eqref{KK-mass} for $m^2_\cK$  based on the conserved $R$ charge,
both sides of \eqref{A-eom-mass-1-loop}
have consistent scaling at late times $\sim e^{-\frac 32 \tau}$ if and only if 
\begin{equation}
\label{eps-34}
\boxed{ \ 
    \varepsilon=-\frac{3}{4} \ . 
    }
\end{equation}
This is precisely the scaling that leads to a constant dilaton and a stable hierarchy, hence the explicit form of $F(\rho)$ is irrelevant.
We will see that this determines the cosmic evolution \eqref{at-modified-NC-1}
\begin{align}
     a(t) \sim \frac 32 t , \qquad 
     H = \frac 1{a}\frac{d a}{dt} \sim \frac 1t
    \label{at-modified-NC-0}
\end{align}
which is quite reasonable for a model without any input from  data; in particular, there is no fine-tuning problem. 
Note that the $\e^{-2\tau}|\dot{f}|^2$ term is negligible at late times, hence the rotating internal dimensions are not significant for the evolution of spacetime.
This is supported by  
the scaling property
\eqref{chi-alpha-34}, which simplifies the computation of the 1-loop effective action for the deformed backgrounds\footnote{See e.g. \cite{Manta:2024vol,Steinacker:2024huv} for loop computations on the undeformed backgrounds.}. 

Quite generally,
the dilaton is expected to be stabilized through quantum effects, since
\begin{align}
    \rho^2 = \frac{\sqrt{|G|}}{\rho_M} \ 
\end{align}
characterizes the density of states per volume; here $\rho_M$ is the symplectic density of quantum spacetime. This characterizes the strength of quantum effects, and an equilibrium with the classical action suggests $\rho = const$.

We have therefore found compelling arguments why quantum effects should stabilize covariant quantum spacetime within the IKKT model such that the dilaton is constant, leading to a specific prediction \eqref{eps-34}, \eqref{at-modified-NC-0} for the evolution of spacetime.

Further details of such quantum corrections and their implications will be studied in future work, refining the preliminary analysis in \cite{Battista:2023glw} taking into account the new results for $m_\cK$.


\section{Formal considerations and  conservation laws}
\label{sec:formal}

\subsection{R-symmetry current conservation}
\label{sec:R-current}

The matrix model enjoys a global $SO(6)$ R-symmetry acting on the 6 internal matrices $T_{\ib}$ as 
$\delta T_{\ib} = \lambda_{\ib}^{\ \jb} T_{\jb}$.
This leads to a conservation law \cite{Polychronakos:2013fma}
\begin{align}
 0 & = \lambda_{\ib \jb} [T_B, \{ T^{\jb}, [T^B,T^{\ib}] \} ] \nn\\
  &=  \lambda_{\ib \jb} [T_{\dot\mu}, \{ T^{\jb}, [T^{\dot\mu},T^{\ib}] \} ]
   +\lambda_{\ib \jb}[T_{\kb}, \{ T^{\jb}, [T^{\kb}, T^{\ib}] \} ] \nn\\
  &= -i[T_{\dot\mu},J^{\dot\mu}]
   + K 
\end{align}
(assuming no fermionic condensate, and $B=0,...,9$)
as a consequence of the classical matrix equations of motion $\square T^{\ib} = 0$.
Here  
\begin{align}
    J^{\dot\mu}
    = i\lambda_{\ib \jb} \{ T^{\jb}, [ T^{\dot\mu},T^{\ib}] \} 
    \end{align}
and 
\begin{align}
\label{K-vanish}
    K = \lambda_{\ib \jb} [T_{\kb}, \{ T^{\jb}, [T^{\kb}, T^{\ib}] \} ]
     &= \lambda_{\ib \jb} \{T^{\jb},\square_\cK T^{\ib} \}  = \L \lambda_{\ib \jb} [T^{\jb}, T^{\ib}]
\end{align} 
assuming $\square_\cK T^{\ib} = \L T^{\ib}$.
 This $SO(6)$ symmetry is spontaneously broken in the presence of $\cK$. 
For the rotating background \eqref{rotating-ansatz}, it is spontaneously broken to some $U(1) \subset SO(6)$ according to $\delta T_{\jb}^\pm = \pm i  T_{\jb}^\pm$  in complex notation.
Then $K$ reduces to
\begin{align}
\label{K-vanish}
   K \sim \sum_{\ib}  [\cK^{\ib +},\cK_{\ib}^-] 
    \stackrel{!}{=} 0
\end{align} 
as in  \eqref{K-condition-eom}, which we assume to vanish.
This condition is also present in the eom \eqref{K-condition-eom}
for $\cM$, and it means that the Yang-Mills gauge current $\cJ_\mu = [T_{\ib},D_\mu T^{\ib}]$ vanishes.
A simple  example of $\cK$ where this condition as well as  \eqref{C-condition} holds is the fuzzy torus; a more interesting one is squashed $\C P^{2}$  \cite{Steinacker:2014eua} and generalizations thereof \cite{Sperling:2018hys}, which features a self-intersecting geometry.
 
In the semi-classical regime, we can rewrite this $R$ current
 using $[T^{\dot\mu},.] \sim i E^{\dot\mu\nu}\partial_\nu$ as 
\begin{align}\label{U1-R-current}
    J^{\dot\mu}
    =  \frac{1}{4}E^{\dot\mu\nu} \sum_{{\jb}=1}^3\big(
    T_{\jb}^- \partial_\nu T_{\jb}^+  +  \partial_\nu T_{\jb}^+ T_{\jb}^-
    - T_{\jb}^+ \partial_\nu T_{\jb}^-
    - \partial_\nu T_{\jb}^-T_{\jb}^+ \big)
     =:  E^{\dot\mu\nu} J_\nu \ 
\end{align}
where $J_\mu$ is recognized as the standard current associated to the global $U(1)\subset SO(6)$ R-symmetry familiar from field theory.
The conservation law can then be written in a covariant form using the Weitzenböck connection \cite{Steinacker:2020xph} associated to the frame $E$, 
noting that 
\begin{align}\label{current-cons-connection}
    \nabla_\mu J^\mu =  \nabla_\mu (E^{\dot\nu\mu} J_{\dot\nu}) =  E^{\dot\nu\mu} \nabla_\mu J_{\dot\nu} = \{T^{\dot\nu},J_{\dot\nu}\} = 0 \ .
\end{align}
Note that this conservation law holds also at the quantum level.

For the present background, the current points along the cosmic time-like vector field, so that it doesn't break any symmetries of the FLRW spacetime.
Assuming the ansatz \eqref{rotating-ansatz},
the current  takes the form
\begin{align}
    J_\mu &=  
    \frac{1}{4}\sum_{{\jb}} \, \big(\bar{f}\partial_\mu f (\cK_{\jb}^- \cK_{\jb}^+
    + \cK_{\jb}^+ \cK_{\jb}^-)
    - f\partial_\mu \bar{f}  (\cK_{\jb}^+ \cK_{\jb}^- + \cK_{\jb}^- \cK_{\jb}^+)\big)  \nn\\
    &= \frac{1}{2}(\bar{f}\partial_\mu f - f\partial_\mu \bar{f}) (\cK_{\ib} \cK^{\ib}) = -iC^2\chi^2\partial_\mu\omega 
\end{align}
for $f = \chi(\tau)e^{i\omega(\tau)}$, or 
equivalently (dropping the overall $C^2$ for simplicity)
\begin{align}
    J^{\dot\mu} &= -i\chi^2 \{T^{\dot\mu},\omega\}
     =: j  x^{\dot\mu} 
\end{align}
with 
\begin{align}
\label{j-explicit}
    j =  \frac{i}{R}\chi^2 \a \omega' \ .
\end{align}
Then current conservation gives 
\begin{align}
0 = \{T^{\dot\nu},J_{\dot\nu}\}
 &= \{T_{\dot\mu},j x^{\dot\mu}\}  
 =  x^{\dot\mu}\{T_{\dot\mu},j \} 
 + j \{T_{\dot\mu},x^{\dot\mu}\}  \nn\\
 &= -\frac{\a}{R} j' x^{\dot\mu}x_{\dot\mu} 
  +  j E_{\dot\mu}^{\dot\mu} \nn\\
 &= R \a j' \cosh^2(\tau) 
  +  j (4\a\sinh(\tau) + R \a' \cosh^2(\tau)) 
\end{align}
hence
\begin{align}
\boxed{
    0 = R(\a j)' \cosh^2(\tau) + 4(\a j) \sinh(\tau) 
    \ . }
\end{align}
This is a first-order linear ODE for $ \a j \sim \chi^2\a^2\omega'$, which captures the rotational part of the motion.
We can easily solve this by
\begin{align}
    \frac{(\a j)'}{\a j}  &= - \frac 4R \frac{\sinh(\tau)}{\cosh^2(\tau)} = - 4 \frac{x_4}{R^2 +x_4^2}
     = -2 \ln(R^2+x_4^2)'
\end{align}
hence
\begin{align}
\a j = \frac{i}{R} \chi^2 \a^2 \omega' = \frac \ell{(R^2+x_4^2)^2}
\end{align}
for some constant $\ell\in \R$, which parametrizes the conserved $R$ charge. This reduces to \eqref{omegadot-J-solution} at late times.

In tensorial language, current conservation thus amounts to
\begin{align}
\nabla^\mu J_\mu = 0 = 
\nabla^\mu(|f|^2 \partial_\mu \omega)
 =  \rho^2 \nabla^{(G) \mu}\big(\chi^2 \partial_{\mu} \omega \big)
\end{align}
where $\nabla$ is the Weitzenböck connection w.r.t. the frame $E$, and $\nabla^{(G)}$ the Levi-Civita connection for the effective metric,
using  \eqref{div-weitze-levi}.



\subsection{Energy-momentum-conservation}
\label{sec:e-m-tensor}

We can now apply a similar analysis to the conservation of the matrix
energy-momentum tensor\footnote{This is an easy consequence of the matrix equations of motion. It can also be derived analogous to Noether's theorem, cf. \cite{Balasin:2015hna, Polychronakos:2013fma, Steinacker:2008ri}.}:
\begin{align}
\label{T-cons}
    [T_A,\cT^{AB}] = 0 \ .
\end{align}
Here
\begin{equation}
\cT^{AB}=[T^A,T^C][T^B,T_C]-\frac{1}{4}\eta^{AB}[T^A,T^C][T_A,T_C]
\end{equation}
is related to the standard energy-momentum tensor 
via the frame as \cite{Steinacker:2024unq}
\begin{align}
    T_{\mu\nu} = \cT^{AB} E_{a\mu} E_{b\nu} 
\end{align}
(possibly up to some power of the dilaton $\rho$). 
This provides another conservation law for the background $\cM$ 
which constrains the dynamics, and which is expected to hold also
at the quantum level.
Note that this includes contributions from the geometry (which is our focus here, but which have no analog in the conventional framework) as well as the standard contributions from matter, which we largely ignore here. 
Hence formulating the dynamics using the conservation law \eqref{T-cons} allows to include also the contributions from matter, and provides some much-needed physical intuition. This should be explored in the future.

In Appendix \ref{em-tensor-computation} we compute (at late times)
\begin{align}
    \mathcal{T}^{\mu\nu}=&-\frac{\eta^{\mu\nu}}{2R^2}\left(Px_4^2+Q-2SR^2\right)+r^2Pt^\mu t^\nu-\left(P+Qx_4^{-2}\right)\frac{x^\mu x^\nu}{R^2}
\end{align}
with
\begin{align}
    P&=\frac{(A+\dot A/2)^2}{r^2R^2}\\
    Q&={A|\dot f|^2C^2}\\
    S&=\frac{|f|^4\Theta^{\ib\jb}\Theta_{\ib\jb}}{4}=\frac{|f|^4\Lambda C^2}{4} \ .
\end{align}
Here $P$ is the spacetime contribution, $S$ can be interpreted as a potential energy of $\cK$, while
$Q$ is proportional to the kinetic energy of $\cK$,
\begin{equation}\label{kin-energy}
    A|\dot f|^2=A(\dot\chi^2+\chi^2\dot\omega^2)=A\dot\chi^2+\frac{\ell^2\e^{-6\tau}}{A\chi^2}
\end{equation}
where we used $R$ charge conservation \eqref{omegadot-J-solution}. 
The corresponding conservation equation is
\begin{equation}\label{cons-em-tensor}
  3P+\frac{\dot{P}}{2}+x_4^{-2}\left(\frac{\dot{Q}}{2}+2Q+\dot{S}R^2\right)=0 \ 
\end{equation}
which can be rewritten as
\begin{equation}
\e^{-4\tau}
\partial_\tau(\e^{6\tau}P)
 = -\dot{S} - \frac 1{2R^{2}}\e^{-4\tau}
\partial_\tau (\e^{4\tau}Q) \ .
\end{equation}
This can be interpreted in terms of energy conservation, where the lhs characterizes the change of energy of spacetime, and the rhs describes the change of the internal energy for $\cK$. 

Now consider the local physics, where cosmological scale variations $\dot \a(\tau)$  are negligible.
Then the above equation reduces to 
\begin{equation}
    \frac{\diff}{\diff\tau}\left(\frac{Q}{2}+R^2S\right)\approx 0
\end{equation}
This means that the internal energy of $\cK$ should be essentially constant, and the contribution from $\omega^2$ should prevent it from shrinking, as classically. More explicitly, assume derivatives of $\e^{-\tau}$ can be neglected at local scales. Then this is a conservation equation for the energy
\begin{align}
    E&=\frac{1}{AC^2}\left(\frac{Q}{2}+R^2S\right)=\frac{\dot\chi^2}{2}+V_{\textnormal{eff}},\ \ \frac{\diff E}{\diff \tau}\approx 0\\
    V_{\textnormal{eff}}&=\frac{\ell^2\e^{-(6+4\varepsilon)\tau}}{2\alpha_0^4\chi^2}+\frac{R^2\Lambda \e^{-2\varepsilon\tau} }{4\alpha_0^2}\chi^4
\end{align}
using $\alpha=\alpha_0\e^{\varepsilon\tau}$, and \eqref{kin-energy}.
The second term in $V_{\textnormal{eff}}$ is the potential energy for $\cK$, while the first term is the contribution from internal angular momentum.

In the non-rotating case $\ell=0$, the effective potential is minimized at $\chi=0$, making the extra dimensions locally unstable.
However in the rotating case $\ell\neq 0$, the potential has a non-trivial minimum, at
\begin{equation}
    \chi^6=\frac{\ell^2 \e^{-(6+2\varepsilon)\tau}}{\alpha_0^2R^2\Lambda} \ .
\end{equation}
This leads to 
\begin{equation}
    \chi \sim \e^{-(1+\frac{1}{3}\varepsilon)\tau}
\end{equation}
which for $\varepsilon = -\frac{3}{4}$ is
$\chi \sim \e^{-\frac{3}{4}\tau}$. This matches precisely the behavior found in \eqref{KK-mass}, in the large angular momentum regime.

\section{Effective frame and metric}\label{eff-frame-metric}

For generic deformations of the background, the effective frame $E^{\dot\alpha \nu}= \{T^{\dot\alpha},x^\nu\}$ acquires $\hs$-valued components.
As shown in \cite{Steinacker:2024unq}, it is always possible to eliminate these by choosing suitable local normal coordinates (LNC) near some reference point $\xi$, in a neighborhood smaller than the local curvature scale. In these local patches, the matrix model action for the 
fluctuation modes
takes the standard form of a kinetic action. By covering spacetime with a collection of local patches, we can thus extract an effective description in terms of an effective (pseudo-) Riemannian metric and transition functions.
In this Section, we illustrate how this works for the present cosmological background.

\subsection{Eliminating $\hs$ in local normal coordinates}

A general $\hs$ valued field on the present spacetime takes the form 
\begin{align}
    \phi = \phi(x) + \phi_{(1)}^\nu(y)u_\nu + \phi_{(2)}^{\nu_1\nu_2}(y)u_{\nu_1}u_{\nu_2}+\ldots .
\end{align}
Here $u^\mu$ are normalized $\hs$ generators 
\begin{align}\label{normalized-u}
    u^\mu = \frac{r}{\cosh\tau} t^\mu, \qquad u_\mu u^\mu = 1 \ 
\end{align}
which generate the $S^2$ fiber of the 6-dimensional bundle space $\C P^{1,2}$ over $\cM^{3,1}$;
they satisfy simple relations given in Appendix  \ref{sec:normalized-u}.

An analogous $\hs$ expansion applies to the frame $E^{\a\mu}$ for the deformed background, which
for generic $\alpha(\tau)$ takes the explicit form
\begin{equation}
   E^{\dot\alpha \nu}= \{\alpha t^{\dot\alpha},x^\nu\}=\alpha\frac{x_4}{R}\eta^{\dot\alpha \nu}+r^2R\alpha^\prime t^{\dot\alpha}t^\nu \ .
\end{equation}
The corresponding effective (auxiliary) metric\footnote{The physical effective metric is derived in Section \ref{dilaton}.} is also $\hs$ valued,
\begin{equation}
    \gamma^{\mu\nu}=\eta_{\dot\alpha \dot\beta}E^{\dot\alpha \mu}E^{\dot\beta \nu}=\alpha^2\frac{x_4^2}{R^2}\eta^{\mu\nu}+r^2\alpha^\prime(2\alpha x_4+R^2\cosh^2\tau \alpha^\prime)t^\mu t^\nu \ .
\end{equation}
It is easy to see that 
for $\alpha\sim \frac{1}{x_4}\sim e^{-\tau}$, both become approximately degenerate for large $\tau$, cf. \eqref{eff-metric-App}.
Therefore this classical solution is not acceptable.

As shown in \cite{Steinacker:2024unq,Battista:2023glw}, the $\hs$-valued components of the frame can be eliminated at any given reference point $\xi$, by choosing appropriate local normal coordinates (LNC) $\tilde x^\mu$. This leads to a standard frame and effective metric, in some sufficiently small neighborhood of $\xi$. This provides the physical interpretation for the background.

Given the manifest $SO(1,3)$ symmetry of the background, 
one might hope to find $SO(1,3)$-covariant 
 Cartesian normal coordinates $\tilde x^\mu$ such that the $\hs$ components of the frame vanish on the entire space-like slice $H^3$. Unfortunately, it turns out that this is not possible.
However, we can find $SO(3)$ covariant local normal coordinates, which cover the full time-like geodesics through the points\footnote{By $SO(1,3)$ covariance we restrict to these points.}  $\xi = (x^0,0,0,0)$ for any $x^0$.

For these  reference point $\xi = (x^0,0,0,0)$,
consider the $SO(3)$ covariant ansatz
\begin{align}
\label{x-tilde-covar}
  \tilde x^\mu  &= x^\mu + \tilde b u^\mu (u_j x^j), 
  \qquad  \tilde x^\mu|_\xi = x^\mu \ .
\end{align}
Note that $u_j x^j$
 vanishes at the reference points $\xi$ but it is not invariant under $SO(1,3)$.
We will determine $\tilde b = \tilde b(\tau)$ such that
the frame has no $\hs$ components at $\xi$.
In these coordinates,
 the frame defined by the background $T^\a = \a t^\a = \tilde \a u^\a$ for $\tilde\a = \frac{\a}r \sinh(\tau)$ is
\begin{align}
  \tilde E^{\a\mu} =
   \{T^\a,\tilde x^\mu\}
    &= \{\tilde \a u^\a,x^\mu + \tilde b u^\mu (u_j x^j)\} \nn\\
    &\approx \tilde \a \{u^\a,x^\mu \} 
    + \tilde b \tilde \a  \{u^\a,u^\mu (u_j x^j)\} 
    +  \tilde \a' \big(\{x_4,x^\mu + \tilde b u^\mu (u_j x^j)\}\big)u^\a  \nn\\
    &+ \tilde b (\eta^{\a j}u_j u^\mu - u^\a u^\mu) 
    + (\varepsilon+1) 
    (1 + \tilde b)u^\a u^\mu\Big)\nn\\
    &\approx\a \sinh(\tau)\Big(\eta^{\a\mu} + ((\varepsilon+1) 
    (1 + \tilde b) -1)u^\a u^\mu\Big) 
\end{align}
 up to corrections of order 
$O(\frac{x_i}{x_4})\hs$ due to \eqref{x4-u-brackets}; here $\hs$ indicates some polynomial in $u$ or order 1.
We assume that $\tilde b$ is slowly varying so that its brackets can be neglected.
Then for
\begin{align}
    \tilde b = \frac{1}{\varepsilon+1} - 1
\end{align}
one finds\footnote{Note that for generic $\alpha(\tau)$, such $\tilde{b}$ depends on $\tau$. However the corresponding corrections to the frame are then suppressed by a factor of order
$\dot{\tilde{b}} O\left(\frac{x_i}{x_4}\right)\hs$.
Thus for well-behaved $\alpha$ and away from $\varepsilon=-1$, the above discussion is unchanged.}
\begin{align}
     \tilde E^{\a\mu} 
    &= \a \sinh(\tau)\big(\eta^{a\mu} + O\left(\frac{x_i}{x_4}\right) \hs\big) \ .
\end{align}
We conclude that in the local normal coordinates $\tilde x^\mu$,
the $\hs$ components of the frame are negligible in a tubular region of $\xi = (x^0,0,0,0)$ with $O(\frac{x_i}{x_4}) \ll 1$, which is set by the space-like cosmic curvature scale.
Observe that the case $\varepsilon = -1$ is singular, which suggests that we need $\varepsilon \neq -1$ for a well-defined the semi-classical regime.

Next, we compute the Poisson brackets of the local normal coordinates, assuming for simplicity the minimal $n=0$ case with $R=r$:
\begin{align}
 \{\tilde x^0,\tilde x^j\} &= 
      \{x^0,x^j - \tilde b x^0 u_0 u^j\}
    \approx r \frac{1}{1+\varepsilon} x^0 u^j  \nn\\
 \{\tilde x^i,\tilde x^j\} &=  
    \{x^i - \tilde b x^0 u_0 u^i, x^j - \tilde b x^0 u_0 u^j\}  
    %
     \sim r (x^i u^j - x^j u^i)
\end{align}
at late times, assuming $\varepsilon \neq -1$. 
This has the same structure as the undeformed background, up to different pre-factors. Hence the local physics is essentially the same as on the undeformed background, and previous results -- including one-loop computations -- carry over immediately. In particular, the uncertainty scale is still of order 
\begin{align}
    L_{\textnormal{NC}}^2 = O(r\xi_0) \ \sim rx^4
\end{align}
in Cartesian LNC,
as long as $\varepsilon \neq -1$. 

We conclude that the classical region where $\hs$ corrections are negligible is much larger than the uncertainty scale, and comparable with the cosmic curvature scale. 

\subsection{Dilaton and effective metric}\label{dilaton}

Now consider the dilaton, which is a scalar degree of freedom determined via the symplectic volume form and the frame as follows \cite{Steinacker:2010rh}:
\begin{align}
   \tilde \rho^2  = \tilde\rho_M^{-1}\sqrt{|G|} = \tilde\rho_M \det \tilde E^{\a\mu} \ .
\end{align}
Here $\tilde\rho_M$ is the reduced symplectic density on $\cM^{3,1}$ in $\tilde x^\mu$ coordinates, which 
relates traces in the matrix model to integrals in the semi-classical regime.
It is obtained by
rewriting the  symplectic volume form $\Omega = \rho_M d^4 x \Omega_u = \tilde\rho_M d^4 \tilde x \Omega_u$ on $\C P^{1,2}$ in terms of the new coordinates
$(\tilde x^\mu,u^j)$
(here $\Omega_u$ is the normalized volume form on the internal $S^2$), using
\begin{align}
    \det\left(\frac{\partial\tilde x^\mu}{\partial x^\nu}\right) &= 1 + \tilde b u^\mu u_\mu
     = 1 + \tilde b  
    = \frac{1}{1 + \varepsilon} \
\end{align}
at late times.
Recalling  $\rho_M = \frac 1{\sinh\tau}$ in Cartesian coordinates $x^\mu$,
the symplectic density on $\cM^{3,1}$ is given by 
\begin{align}
    \rho_M d^4 x 
   = \tilde \rho_M d^4 \tilde x, \qquad  \tilde \rho_M = \frac{1 + \varepsilon}{\sinh\tau} 
\end{align}
in the $\tilde x^\mu$ coordinates.
Therefore the dilaton is
\begin{align}
    \tilde\rho^2  = \tilde\rho_M \det \tilde E^{\a\mu}
     = (1 + \varepsilon)(\sinh\tau)^3 \a^4 \ ;
     \label{tilde-rho}
\end{align}
the tilde will be dropped in the following.
This is generically not constant during the cosmic evolution, e.g. $\rho \sim \e^{\frac{3}{2}\tau} \sim {a(t)}$ for $\a = 1$. 
However it is constant
at late times for $\a \sim e^{-\frac 34 \tau}$
\begin{align}
\label{epsilon-dilaton-const}
   \rho  = const \qquad \mbox{for} \quad \varepsilon = -\frac{3}{4} \ .
\end{align}
This is desirable on physical grounds, since then the gauge couplings $g^2_{YM}\propto \rho^{-2}$ \cite{Steinacker:2024huv}  would be independent of time, as well as other physical observables such as the density of states, KK scales, etc. More generally, the validity of the weak coupling regime requires $\varepsilon\geq -\frac{3}{4}$ at late times. The case $\varepsilon=-\frac{3}{4}$ appears to be dynamically preferred as argued in Section \ref{asymptotic-sols}. 

The effective metric in a tubular region around the time-like geodesics $\xi = (\xi^0,0,0,0)$ in local normal coordinates
is therefore given by 
\begin{align}
G^{\mu\nu} &= \rho^{-2} \eta_{AB}E^{A\mu} E^{B\nu} 
 = \frac 1{\a^2\sinh \tau (1 + \varepsilon)} \eta^{\mu\nu},   \
\label{eff-metric-App} 
\end{align}
up to corrections of order $O\big(\frac{\tilde x^i}{\tilde x^0} \hs\big)$.
While the $SO(1,3)$ symmetry is no longer manifest in these  coordinates, the background and therefore the physics  certainly are invariant. Therefore we can match this with a unique $k=-1$ FLRW metric. 
Using the hyperbolic parametrization \eqref{embedding-3d-hyperboloid} within this tubular region for the $\tilde x^\mu$,
we can write 
the effective metric near $\xi$ 
as
\begin{align}
d s^2_G &= (1 + \varepsilon)\a^{2}\sinh \tau \,  \eta_{\mu\nu} d\tilde x^\mu d\tilde x^\nu \nn\\
&\approx   (1 + \varepsilon) \a^{2} \sinh \tau R^2\big(- \sinh^2\tau \,   d \tau^2 
  +  \cosh^2\tau\, d \Sigma^2 \big) \stackrel{!}{=} -d t^2 + a(t)^2 d\Sigma^2
\label{eff-G-FLRW}
\end{align}
 where 
\begin{align}
    d\Sigma^2 = d\chi^2 + \sinh^2\chi (d\theta^2 + \sin^2 \theta \, d\varphi^2)
    \label{dSigma2}
\end{align}
is the invariant metric on the space-like 3-hyperboloids $H^3$.
This is recognized as a $k=-1$ FLRW metric:
\begin{subequations}
\begin{align}
dt^2 &=  R^2  (1 + \varepsilon)\alpha^2 \sinh^3 \tau \, d \tau^2 
\approx R^2  (1 + \varepsilon) \alpha^2 e^{3\tau} \, d \tau^2 ,  
\\
a^2(t) &= R^2  (1 + \varepsilon) \alpha^2 \sinh\tau\cosh^2\tau  
\approx R^2  (1 + \varepsilon) \alpha^2 e^{3\tau}
\end{align}    
\label{dt-dtau-a-relation}
\end{subequations}
at late times, with FLRW time parameter
\begin{align}\label{t-tau-rel-cosm}
    t(\tau) = R\int_0^\tau  \sqrt{1+\varepsilon} \a e^{\frac{3}{2}\tau} d\tau
\end{align}
where
$\varepsilon= \a^{-1}\frac{d\a}{d\tau}$  \eqref{vareps-def}. For $\a = \a_0 e^{\varepsilon\tau}$, this gives 
\begin{align}
    t \sim \a_0 R  \frac{\sqrt{1+\varepsilon}}{\varepsilon + \frac 32}\, e^{(\varepsilon + \frac 32)\tau}
\end{align}
 and the cosmic scale parameter is obtained as
\begin{align}
     a(t) \sim  R \sqrt{(1 + \varepsilon)} \alpha e^{\frac{3}{2}\tau}
     \sim \a_0 R\sqrt{(1 + \varepsilon)}  e^{(\varepsilon + \frac 32) \tau}
     \sim (\varepsilon + \frac 32) t
    \label{at-modified-NC-1}
\end{align}
at late times, corresponding to an expanding asymptotically "coasting" FLRW cosmology.
In particular, the Hubble parameter is found to be
\begin{align}
    H = \frac 1{a(t)}\frac{d }{dt} a(t)
    = 
       \frac 1t
\end{align}
which is quite reasonable for a model without any input\footnote{It is expected that matter will have some impact on the cosmic evolution, but far less than in GR.}.
Similar results are obtained 
for $k=0$ cosmological quantum spacetime in a forthcoming paper.

We also note that the  classical solution $\varepsilon=-\frac{3}{2}$ leads to $a(t)=const$ i.e. Minkowski metric, while the dilaton decays rapidly. We therefore discard this solution as unphysical.

\subsection{Higher-spin manifold and local structure}

We have seen that the deformed background leads to local normal coordinates around any point $\xi$ on $\cM^{3,1}$, such that the $\hs$ components 
are negligible in the locally flat regime.
In this Section 
we show that for nearby $\xi$, the algebras generated by these local normal coordinates coincide.
This means that the local tensor fields around sufficiently close reference points are related by the usual concept of a transition function, 
with negligible $\hs$ corrections.
This suggests the concept of a higher-spin manifold which behaves like a standard manifold for local scalar functions, 
which is equipped locally with a standard notion of frame, metric and a dilaton. We use the present setting to discuss some aspects of such a framework.

\paragraph{Relating coordinate patches.}

Consider the deformed  background with local normal coordinates defined by \eqref{x-tilde-covar} around two reference points 
$\xi^\mu= (\xi^0,\xi^i)$ and ${\xi'}^\mu  = \L^\mu_{\ \nu} \xi^\nu \approx (\xi'^0,0)$,   related by a  translation $\L^\mu_{\ \nu} \in SO(1,3)$ along the space-like $H^3$ within the locally flat regime,
such that ${x'}^j|_{\xi'} = 0$. Then 
 $x'^\mu = \L^\mu_{\ \nu} x^\nu$, while 
$u'^\mu = \L^\mu_{\ \nu} u^\nu \approx u^\mu$
since $u^0 \approx 0$ near the reference point. 
The associated local normal coordinates $\tilde x^\mu$ and $\tilde x'^\mu$  are hence given by
\begin{align}
  \tilde {x'}^\mu  &=  {x'}^\mu + \tilde b (u'_j {x'}^j)\, {u'}^\mu \approx  {x'}^\mu + \tilde b (u_j {x'}^j)\, {u}^\mu  \nn\\
\tilde {x}^\mu  &=  {x}^\mu + \tilde b (u_j {x}^j)\, {u}^\mu \ .
\end{align}
For small translations in this neighborhood, the time-like coordinates satisfy
$\tilde {x'}^0 \approx x'^0 \approx  x^0$. 
The space-like components can be related by 
 $x'^i \approx x^i - \xi^i$, so that
\begin{align}
  \tilde {x'}^i 
   &\approx {x}^i - \xi^i
   + \tilde b  u_j ({x}^j - \xi^j)\, {u}^i  \
    =: \tilde {x}^i - \tilde \xi^i
\end{align}
and hence  
\begin{align}
\boxed{
  \tilde {x'}^i 
   \approx  \tilde {x}^i - \tilde \xi^i \ ,
   \quad \tilde  {x'}^0  \approx  \tilde {x}^0 .
   }
\end{align}
We define the algebra 
\begin{align}
   \widetilde{\cC}^0 := \C[[\tilde x^\mu]] 
\end{align}
of functions (power series) generated by the local normal coordinates $\tilde x^\mu$. Then the above result implies that these local algebras $\widetilde{\cC}^0$ essentially coincide
for all points within the locally flat regime, and can be identified with a commutative algebra of functions 
near $\xi \in \cM^{3,1}$.
We can 
 thus compute the transition functions of local tensor fields in a standard manner.
In particular, (local) scalar fields are elements  $\phi \in \widetilde{\cC}^0$.
By construction, the frames $\tilde E^{a\mu}$ are also elements in $\widetilde{\cC}^0$, and hence define an effective $(3+1)$-dimensional metric 
with Minkowski signature
for local scalar functions (and tensor fields).

More geometrically, local normal coordinates $\tilde x^\mu$ can be viewed as  maps from the 6-dimensional bundle space to a $(3+1)$-dimensional base space $\widetilde{\cM}^{3,1}$, which plays the role of spacetime for the given background. 
The $\tilde x^\mu$ can alternatively be viewed as
$\hs$-valued functions on the undeformed base $\cM^{3,1}$.
The points $\xi$ are always defined\footnote{We assume here that the deformation is sufficiently mild.} in terms of the undeformed $\cM$, and we will always use the same generators $u^\mu$ for the internal $S^2$ fiber.
This is our working definition of a $\hs$ manifold.

To make this more explicit,
consider a
scalar field $\phi(\tilde x) = e^{i k \tilde x}$ in some local normal coordinate patch. Its  expression in a distant patch may in general involve non-trivial $\hs$ components.
However for a shifted $\tilde x'$ as above, it reads
\begin{align}
    e^{i k_\mu \tilde x^\mu}
    &= e^{i k_\mu(\tilde x'^\mu + \tilde\xi^\mu)}
     =  e^{i k_\mu \tilde x'^\mu} e^{i k_\mu \tilde \xi^\mu} \ .
\end{align}
Even though the second factor is globally $\hs$-valued, it is locally $\widetilde{\cC}^0$-valued.
Hence the transition functions are "truly" $\hs$ valued only if we go beyond the locally flat regime defined by the cosmic curvature.
This  means that the usual concepts of tensor fields 
on standard manifolds carry over to the present setting, within locally flat coordinate patches. 

At larger distances, a free propagating  scalar field is expected to develop $\hs$ components. 
This would be avoided if the $\hs$ modes acquire a mass via quantum effects. That issue remains to be studied in future work.

\section{Discussion}
In this work we have constructed and studied a class of $SO(1,3)$ covariant cosmological quantum spacetimes $\mathcal{M}^{3,1}$ with time-dependent extra dimensions $\cK$, as classical solutions for the supersymmetric IKKT matrix model \cite{Ishibashi:1996xs}. This is a significant advancement over previous constructions, which required explicit mass deformations breaking the model's fundamental supersymmetry \cite{Sperling:2018xrm}.

Our key result is the identification of a robust classical stabilization mechanism for the fuzzy extra dimensions $\cK$. By giving them an non-vanishing internal angular momentum, identified with a conserved R charge (from the global $SO(6)$ symmetry of the model), we have shown that $\cK$ can be prevented from collapsing, leading to a stable minimum for the scale $\chi(\tau)$ of the extra dimensions. This mechanism naturally generates  and stabilizes a large hierarchy between the UV scale $m_\cK\sim \chi$ and the IR scale of the cosmological background $m_{\textnormal{cosm}}\sim \frac{\alpha(\tau)}{R}$, which is essential for realistic effective physics.

We have also demonstrated that the coupled equations of motion for the spacetime scale factor $\alpha(\tau)$ and the internal scale $f(\tau)=\chi(\tau)\e^{i\omega(\tau)}$ can be recast as conservation laws for the R current and the matrix energy-momentum tensor. This reformulation offers a transparent and physically intuitive picture of the dynamics.

The classical dynamics can be solved for both $\alpha(\tau)$ and $f(\tau)$. The solution for spacetime is found to  $\alpha(\tau)\sim \e^{-\tau}$ at late times, consistent with \cite{Battista:2023glw} and quite independent of $\cK$. This solution is however physically problematic, as it leads to decreasing dilaton and thus strongly coupled Yang-Mills gauge theory at late times, invalidating the semi-classical approximation. Moreover, the metric becomes (almost-) degenerate at late times.
We argue that quantum effects resolve this issue, and a self-consistent treatment including quantum effects
uniquely selects the scaling $\alpha(\tau)\sim \e^{-\frac{3}{4}\tau}$. This critical scaling leads to constant dilaton $\rho$ at late times, ensuring that the YM coupling and other physical parameters remain constant throughout late-time cosmic evolution. Furthermore, this results in a linearly expanding scale factor $a(t)\sim t$, with Hubble parameter $H(t)= \frac{1}{t}$, providing a simple yet physically reasonable cosmological history, given the simple ansatz without taking into account matter.

This picture suggests a balance where classical dynamics provides the mechanism for stabilizing the extra dimensions, while quantum effects stabilize the dilaton and determine the precise dynamics of spacetime. The resulting background, with stable hierarchy and constant couplings, provides a consistent and promising vacuum for the IKKT model.

The effective geometry on this background is governed by a higher-spin noncommutative gauge theory, as the fluctuations take values in the higher-spin algebra associated to  covariant quantum spacetimes. We have outlined the resulting concept of higher-spin manifold, showing that local physics can be described by standard Riemannian geometry by employing appropriate local normal coordinates. 

\paragraph{Outlook.}
This work opens up several directions. The first and foremost task is to fully incorporate quantum effects including vacuum energy at least at one loop, refining previous works \cite{Steinacker:2024huv, Manta:2024vol,Battista:2023glw}. This would allow to verify the arguments for the spacetime evolution $\alpha(\tau)\sim \e^{-\frac{3}{4}\tau}$
at the quantum level.
The one-loop effective action contains rich gauge and gravitational dynamics, which should be compared with known physics. In particular, new insights related to "dark energy" can be expected. This should also allow a more explicit characterization of the internal space $\cK$.

Another natural direction is the 
inclusion of matter through the 
 energy-momentum tensor, extending the treatment in Section \ref{sec:e-m-tensor}. This is clearly 
 crucial in addressing the detailed cosmological dynamics, both at a classical and quantum level.
 A numerical analysis of the coupled system $\mathcal{M}^{3,1}\times_\tau \cK$ -- including quantum fluctuations -- would be desirable to establish the dynamical stability of the proposed solution.
 
Finally, the role of the higher-spin modes is not yet fully understood. It can be expected (or hoped) that most of them acquire a mass through quantum effects. A detailed study of their effects on cosmological scales and their decoupling at low energy is required to understand this issue.

\subsection*{Acknowledgements}

We would like to thank Christian Gaß for useful discussions, and Satoshi Iso for pointing out ref. \cite{Iso:2015mva}.
This work is supported by the Austrian Science Fund (FWF) grant P36479.

\appendix

\section{Notation}\label{notation}
Throughout the work we use the following symbols:
\begin{itemize}
    \item $g$: matrix model coupling.
    \item $\mathcal{M}^{3,1}\equiv\mathcal{M}^{3,1}_n$: covariant cosmological quantum spacetime.
    \item $\mathcal{K}$: fuzzy compact extra dimensions.
    \item $\mathfrak{hs}$: algebra or sector of higher spin valued functions.
    \item $n$: quantum number labeling $SO(4,2)$ representation $\mathcal{H}_n$ underlying $\mathcal{M}^{3,1}_n$.
    \item $r$: fundamental length scale of matrices $X^a$ in $\mathcal{M}^{3,1}$.
    \item $R=R_n=\frac{rn}{2}$: radius of 4-hyperboloid underlying $\mathcal{M}^{3,1}_n$ \eqref{x-R-constraint}. 
    
    \item $\square_T=[T^a,[T_a,\ \cdot \ ]]$: matrix d'Alembertian/Laplacian.
    \item $\Lambda$: eigenvalue of $\square_\cK \cK^{\ib}=\Lambda \cK^{\ib}$.
    \item $C^2$: radial constraint $\cK^{\ib}\cK_{\ib}=C^2\one$.
    \item $m_{\cK}$: Kaluza-Klein mass scale for $\cK$.
    \item $\tau$: $SO(1,3)$ invariant cosmological time defined via
    $x_4=R\sinh\tau$.
    \item $(\cdot)'=\frac{\partial}{\partial x_4}$
    \item $\dot{\alpha}=\frac{\partial \alpha}{\partial \tau}$.
    \item $\varepsilon = \a^{-1} \dot \a$: cosmological variation rate of spacetime scale.
    \item $\ell$: $U(1)\subset SO(6)$ R-charge (internal angular momentum).
\end{itemize}
and the following abbreviations:
\begin{itemize}
    \item FLRW: Friedmann–Lemaître–Robertson–Walker. 
    \item Eom: equations of motion.
    \item LNC: local normal coordinates.
    \item KK: Kaluza-Klein.
    \item NC: noncommutative.
    \item YM: Yang-Mills.
    \item Lhs, rhs: left/right hand side.
\end{itemize}
\section{Useful Identities}\label{basic-identities}
We work in the semi-classical regime, where commutators of quantized functions over the symplectic spaces $\mathcal{M}^{3,1}$ can be identified with Poisson brackets
\begin{equation}
    [\ \cdot \,\ \cdot \ ]\sim i\{\ \cdot \ ,\ \cdot \ \} \ .
\end{equation}
We use the following basic semi-classical identities for the background under consideration
\begin{align}
    x^\mu x_\mu &=-R^2-x_4^2\\
    t^\mu t_\mu &=\frac{R^2+x_4^2}{r^2R^2} \label{x-R-constraint}  \\
    t^\mu x_\mu &=0\\
    \{t^\mu, x^\mu\} &=\frac{x_4}{R}\eta^{\mu\nu}\\
    \{x^\mu,x^\nu\} &=\theta^{\mu\nu}=-r^2R^2\{t^\mu,t^\nu\}\\
    \{\theta^{\mu\nu},x^\sigma\} &=-r^2(\eta^{\mu\sigma}x^\nu-\eta^{\nu\sigma}x^\mu)\\
    \{\theta^{\mu\nu},t^\sigma\} &=-r^2(\eta^{\mu\sigma}t^\nu-\eta^{\nu\sigma}t^\mu)\\
    \{t^\mu,x_4\} &=-\frac{1}{R}x^\mu, \quad\{x^\mu,x_4\}=-r^2Rt^\mu\\
    \theta^{\mu\nu} &=\frac{r^2R}{R^2+x_4^2}(x_4(x^\mu t^\nu-x^\nu t^\mu)+\varepsilon^{\mu\nu\rho\sigma}x_\rho t_\sigma)
\end{align}
with $\theta^{\mu\nu}$ generators of the $SO(1,3)$ algebra of the doubleton representation from which the matrices are constructed, and $R \sim \frac n2 r$. Note that the $\varepsilon^{\mu\nu\rho\sigma}$ term in $\theta^{\mu\nu}$ is not present in the minimal case $n=0$ \cite{Manta:2025inq}. We mostly work in the late time regime $x_4\gg R$.

\section{Normalized $\hs$ generators}
\label{sec:normalized-u}

Consider the rescaled generators
     $\a t^\mu$,
which satisfy the brackets 
\begin{align}
\{\a t^\sigma, \a t^\mu \}
   =& -\a^2\Big(\frac{1}{r^2R^2}\theta^{\sigma\mu}
   +\frac 1R\frac{\a^\prime}{\a}(x^\sigma t^\mu
   -t^\sigma x^\mu)
   \Big) \ .
\end{align}
In particular, for $\a = \frac{r}{\cosh(\tau)}$ we obtain the normalized generators 
\begin{align}
     u^\mu = \frac{r}{\cosh(\tau)} t^\mu, \qquad 
     u_\mu u^\mu = 1 \ 
\end{align}
of the internal $S^2$,
which satisfy 
\begin{align}
\{u^\sigma, u^\mu \}
   &=  -\frac{r^2}{R^2\cosh^2(\tau)}\Big(\frac{1}{r^2}\theta^{\sigma\mu}
   -\frac{\sinh(\tau)}{\cosh^2(\tau)}(x^\sigma t^\mu
   -t^\sigma x^\mu)\Big) \nn\\
    & = -\frac{r}{R^2}\frac{1}{\cosh^3(\tau)} \varepsilon^{\sigma\mu\rho\nu} x_\rho u_\nu \ 
    = O\left(\frac{rR}{x_4^2}\right)
\end{align}   
(which should be considered as approximately zero at late times) and 
\begin{align}
\label{x4-u-brackets}
    \{x^4, u^\mu \}
   &=  \frac{r}{\cosh(\tau)}\{x^4, t^\mu \} = \frac{r}{R\cosh\tau} x^\mu \ = O(r) \nn\\
   \{f(x^4), u^\mu \}
   &=  O(f' r) = O(\frac{f'x_4}{f} \frac{f}{x_4} r)
\end{align}
while 
\begin{align}
    \{u^\mu,x^\nu\} &= r\{\frac{1}{\cosh(\tau)} t^\mu,x^\nu\} 
    = \frac{r x_4}{R\cosh(\tau)} \eta^{\mu\nu} + t^\mu r\{\frac{1}{\cosh(\tau)},x^\nu\} \nn\\
    &\sim r (\eta^{\mu\nu} - u^\mu u^\nu)
\end{align}
for large $\tau$.
The $u^\mu$ brackets are reminiscent of those of a space-like fuzzy sphere, however
the radius is inconsistent with that of a corresponding fuzzy sphere.
For most purposes we can consider the $u^\mu$ as effectively {\bf commuting} variables, and for the minimal case with $n=0$, these brackets would vanish identically\footnote{Note that vanishing here is understood up to corrections smaller than the noncommutativity scale, i.e. negligible in the semiclassical regime. This is compatible with a 6-dimensional symplectic structure because the $u^\mu$ satisfy the constraint $u_\mu u^\mu = 1$.},
\begin{align}
    \{u^\mu, u^\nu \} = 0 \qquad \mbox{for} \ \ n=0  \ .
\end{align}


\section{Computation of the equations of motion}\label{eom-computation}
\paragraph*{Eom for $\cK$.}
The equations of motion for ${\bf{T}}^{\ib}$ can be decomposed into spacetime and internal contributions
\begin{equation}
    \square_{1,9}{\bf{T}^{\ib}}=(\square_{1,3}+\square_6){\bf{T}^{\ib}} \ .
\end{equation}
We can rewrite the internal contribution $\square_{6} $ using complex notation \eqref{complex-TI-def}  as 
\begin{align}
   \square_{6} &=\comm{T_{\ib}}{\comm{T_{\ib}}{\cdot}}
    =\frac{1}{2}(\comm{T_{\ib}^+}{\comm{T_{\ib}^-}{\cdot }}+\comm{T_{\ib}^-}{\comm{T_{\ib}^+}{\cdot }})
\end{align}
so that  we obtain for ${\bf{T}^{\ib +}} = {f}\cK^{\ib +}$
\begin{equation}
    \square_6(f\cK^{\ib +})=(|f|^2\Lambda)(f\cK^{\ib +}) \ .
\end{equation}
where $\Lambda$ is defined in \ref{def-Lambda}.

We will work in the semi-classical regime for the quantized symplectic space $\mathcal{M}^{3,1}$, so that we can work with Poisson brackets.
Then the $\square_{1,3}f$ term gives\footnote{recall $(\cdot)^\prime=\frac{\diff}{\diff x_4}$.}
\begin{align}
    \eta_{\mu\nu}[T^\mu,[T_\mu,f]]=&-\eta_{\mu\nu}\{\alpha t^\mu,\{\alpha t^\nu,f \}\}=\eta_{\mu\nu}\frac{1}{R}\{\alpha t^\mu,\alpha f^\prime x^\nu\}\nn\\
    =&\frac{\alpha}{R^2}\left[\alpha (x_4^2+R^2)\partial_{x_4}^2+x_4(4\alpha +2x_4\alpha^\prime )\partial_{x_4}+R^2\alpha^\prime \partial_{x_4}\right]f \nn\\
    \approx &\frac{1}{R^2}\left[x_4^2\alpha^2 \partial_{x_4}^2+x_4\alpha \left(4\alpha +2x_4\alpha^\prime \right)\partial_{x_4}\right]f 
\end{align}
where the last step holds at late times. Combining these terms,
the eom for ${\bf{T}}^{\ib}$ becomes
\begin{equation}
    \left[x_4^2\alpha^2 \partial_{x_4}^2+x_4\alpha \left(4\alpha +2x_4\alpha^\prime \right)\partial_{x_4}+R^2\Lambda |f|^2 \right]f =0\ .
\end{equation}
With $A=\alpha^2$ we can rewrite this as
\begin{equation}
    {(x_4^2+R^2)Af^{\prime\prime}+4Ax_4f^\prime+(x_4^2+R^2)A^\prime f^\prime+R^2\Lambda |f|^2f=0}
\end{equation}
which simplifies at late times to
\begin{equation}\label{eomKcx}
    {{x_4^2Af^{\prime\prime}+4Ax_4f^\prime+x_4^2A^\prime f^\prime+R^2\Lambda |f|^2f=0}}\ .
\end{equation}

\paragraph{Eom for $\cM$.}
Let us now turn to the eom for 
${\bf{T}}^\mu = \a t^\mu$ defining spacetime:
\begin{equation}\label{eom-M}
    \square_{1,9}{\bf{T}}^\mu=(\square_{1,3}+\square_6){\bf{T}}^\mu= (\square_{1,3}+\square_6)(\alpha t^\mu)\ .
\end{equation}
The first term is
\begin{align}
    \square_{1,3}(\alpha t^\mu)=&-\eta_{\rho\sigma}\{\alpha t^\rho,\{\alpha t^\sigma,\alpha t^\mu\}\}\ .
\end{align}
Let us first compute

\begin{align}
   & \{\alpha t^\sigma,\alpha t^\mu\}=-\frac{1}{r^2R^2}\alpha^2 \theta^{\sigma\mu}+\alpha \frac{\alpha^\prime }{R}t^\sigma x^\mu-\alpha \frac{\alpha^\prime }{R}x^\sigma t^\mu\ .
\end{align}
Then the first term contributes
\begin{align}
    &\frac{1}{r^2R^2}\eta_{\rho\sigma}\{\alpha t^\rho,\alpha^2 \theta^{\sigma\mu}\}=\frac{3\alpha^3 }{R^2}t^\mu-\frac{2\alpha^\prime \alpha^2 }{r^2R^3} \eta_{\rho\sigma}x^\rho\theta^{\sigma\mu}=\frac{1}{R^2}\left(3\alpha^3 +2\alpha^2 \alpha^\prime x_4\right)t^\mu\ .
\end{align}
We compute 
\begin{align}
    &-\frac{1}{R}\eta_{\rho\sigma}\{\alpha t^\rho,\alpha \alpha^\prime t^\sigma x^\mu\}
    =-\frac{1}{R^2}\left(x_4\alpha^2 \alpha^\prime +\alpha (\alpha^\prime )^2(R^2+x_4^2)\right)t^\mu\nn\\
    \approx &-\frac{1}{R^2}\left(x_4\alpha^2 \alpha^\prime +\alpha (\alpha^\prime )^2x_4^2\right)t^\mu\ .
\end{align}
The last term is
\begin{align}
  &\frac{1}{R}  \{\alpha t^\rho,\alpha \alpha^\prime x^\sigma t^\mu\}\nn\\
  &= \frac{\alpha^2 \alpha^\prime }{R^2}(4x_4+R\sinh\tau )t^\mu+\frac{(\alpha^\prime )^2\alpha }{R^2}(x_4^2+R^2)t^\mu +(x_4^2+R^2)\frac{\alpha }{R^2}((\alpha^\prime )^2+\alpha^{\prime\prime} \alpha )t^\mu \nn\\
  &\approx \frac{\alpha }{R^2}\left(\alpha \alpha^\prime (4x_4+R\sinh\tau)+2(\alpha^\prime )^2x_4^2+\alpha^{\prime\prime} \alpha x_4^2\right)t^\mu
\end{align}
leading to
\begin{align}
    \square_{1,3}T^\mu\ 
    &=\frac{\alpha}{R^2}(3\alpha^2+6\alpha\alpha^\prime x_4+(\alpha^\prime)^2(x_4^2+R^2)+\alpha^{\prime\prime}\alpha(x_4^2+R^2) )t^\mu \nn \\
    &\approx  \frac{\alpha }{R^2}\left(3\alpha^2 +6\alpha^\prime \alpha x_4+(\alpha^\prime )^2x_4^2+\alpha^{\prime\prime} \alpha x_4^2\right)t^\mu \ .
\end{align}
The contribution form the internal Laplacian is
\begin{align}
    &\frac{1}{2}[\bar{f}  \cK_{\ib }^-,[f \cK^{\ib +},\alpha t^\mu] ]-\frac{1}{2}[f \cK^{\ib +},[\bar{f}  \cK^-_{\ib},\alpha t^\mu]]\nn\\
    &= -\frac{1}{2}\{\bar{f} ,f^\prime \alpha \frac{x^\mu}{R}\}(\cK^{{\ib} -}\cK_{\ib}^{+}+h.c.)+\frac{1}{2}\frac{\alpha x^\mu}{R}([\cK^{\ib -},\cK_{\ib}^+]\bar{f}f^\prime+h.c.)\nn\\
    &= -r^2|f^\prime|^2C^2\alpha t^\mu+\frac{1}{2}\frac{\alpha }{R}[\cK^{\ib -},\cK_{\ib}^+](\bar{f}f^\prime-f\bar{f}^\prime)x^\mu
\end{align}
with $C^2=\cK^{\ib}\cK_{\ib}>0$. Hence  the exact equations of motion without late time approximation are
\begin{equation}
    \frac{\alpha}{R^2}(3\alpha^2+6\alpha^\prime \alpha x_4+(\alpha^\prime)^2x_4^2+\alpha^{\prime\prime}\alpha x_4^2-C^2r^2R^2|f^\prime|^2+R^2((\alpha^\prime)^2+\alpha^{\prime\prime}\alpha))=0
\end{equation}
and
\begin{equation}
\label{K-condition-eom}
    \boxed{[\cK^{\ib -},\cK_{\ib}^+](\bar{f}f^\prime-f\bar{f}^\prime)=0}\ .
\end{equation}
This suggests that either $f$ is real, which is discussed in Appendix \ref{nonrotatingsol}, or
\begin{align}
\label{K-condition-B}
    K \equiv [\cK^{\ib -},\cK_{\ib}^+]  = 0
\end{align}
in accordance with \eqref{K-vanish},
which we will assume for the backgrounds under consideration\footnote{This holds e.g. for the fuzzy torus,
minimal squashed $\C P^2_N$ \cite{Steinacker:2014lma}  and generalizations thereof \cite{Sperling:2018hys}.}.
We can rewrite the eom in terms of $A=\alpha^2$
\begin{equation}
   3A+3A^\prime x_4+\frac{1}{2}A^{\prime\prime}x_4^2+\frac{1}{2}R^2A^{\prime\prime}-C^2r^2R^2|f^\prime|^2=0\ .
\end{equation}
At late times, 
it is useful to replace $x_4$ by the time coordinate $\tau$ defined via $x_4=R\e^\tau$.
Then the eom for $T^\mu$ eom becomes
\begin{equation}\label{eom-Tmu-tau}
    \boxed{3A+\frac{5}{2}\dot{A}+\frac{1}{2}\ddot{A}-r^2\e^{-2\tau}|\dot{f}|^2C^2=0}
\end{equation}
and the eom for $\cK$ eom becomes
\begin{align}\label{eom-K-tau}
    &\boxed{A\ddot{f}+3A\dot{f}+\dot{A}\dot{f}+R^2|f|^2f\Lambda=0}\\
    = &\ \partial_\tau{(A \dot{f})}+3(A\dot{f})+R^2|f|^2f\Lambda\ .
\end{align}

\section{Oscillatory Solution: real $f$}\label{nonrotatingsol}

If we do not require $K=0$ in \eqref{K-condition-eom}, we can still solve it by requiring $f$ to be real, up to a constant phase. In this Section we will show why this case is physically problematic, and a rotating solution (i.e. dynamical phase for $f$) is appropriate in the present context.\\

Due to the $\e^{-2\tau}$ factor in \eqref{eom-Tmu-tau}, we expect the solution \eqref{alpha-for-static-f}
to dominate the full problem at sufficiently late times, as confirmed numerically.

\begin{figure}[h]
    \centering
    \includegraphics[width=0.8\linewidth]{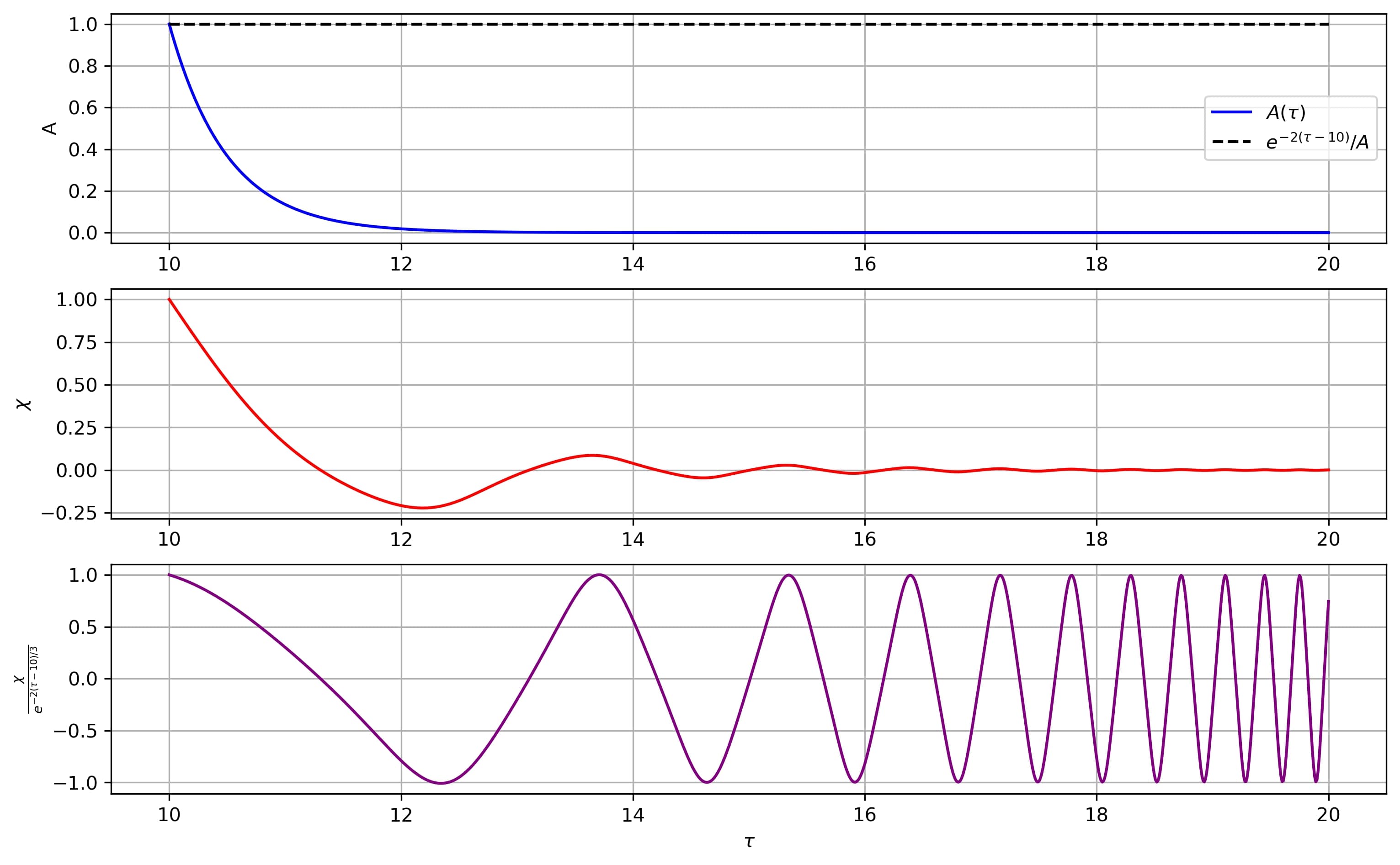}
    \caption{Numerical solutions for non-rotating $f(\tau)$ on the background with $\alpha\sim x_4^{-1}$.}
    \label{non-rotating-num}
\end{figure}

The equation of motion for the scale of the extra dimensions $f$, for $\alpha=\alpha_0\e^{-\tau}$, becomes
\begin{equation}
    \ddot{f}+\dot{f}+\frac{\Lambda R^2}{\alpha_0^2}\e^{2\tau}f^3=0 \ .
\end{equation}
Since $m_\cK^2 = f^2$ should be a UV scale,
we need to consider the non-linear regime where the non-linear term is large.
Writing
\begin{equation}
    f(\tau)=\phi(\tau)\e^{-\tau}
\end{equation}
this becomes
\begin{equation}
    \ddot\phi-\dot\phi=-\frac{\Lambda R^2}{\alpha_0^2}\phi^3\ .
\end{equation}
The lhs can match the rhs only if the time derivatives are very large $\phi$, so that $\ddot\phi\gg \dot\phi\gg \phi$, hence we can approximate
\begin{equation}
    \ddot\phi\approx -\frac{\Lambda R^2}{\alpha_0^2}\phi^3 \ .
\end{equation}
This is solved by a Jacobi Elliptic sine function $\operatorname{sn}$,
\begin{equation}
    \phi\approx f_0 \operatorname{sn}\left(\frac{f_0\sqrt{\Lambda R^2}}{\alpha_0}\tau\biggl.\biggr|-1\right)
\end{equation}
which is highly oscillatory, provided
\begin{equation}
    f_0\gg \frac{\alpha_0}{\sqrt{\Lambda} R}\ .
\end{equation}
This equation is solved numerically in Figure \ref{non-rotating-num}, confirming the  expectations. It gives a rapidly oscillatory solution for $\cK$, with amplitude rapidly decreasing in time, and eventually approaching a regime where the cubic term is suppressed. Then the solution would reduce to the approximate solution $f=f_0\e^{-\tau}$ which varies only at cosmic time scales, provided 
\begin{equation}
    f_0\ll \frac{\alpha_0}{\sqrt{\Lambda} R}\ .
\end{equation}
However the KK masses $m_\cK \sim f$ are then in the IR regime, since for $\alpha=\alpha_0\e^{-\tau}$
\begin{equation}
    \frac{m_{\cK}^2}{m^2_{\textnormal{cosm}}}\sim \frac{1}{\Lambda}
\end{equation}
so that we do not get a large separation of scales required for realistic 4d physics.

\section{Energy-momentum tensor}\label{em-tensor-computation}

We want to compute the matrix energy-momentum tensor
\begin{align}
     \cT^{\mu\nu} &= [T^\mu,T^\sigma][T^\nu,T_\sigma] + [T^\mu,T^{\ib}][T^\nu,T_{\ib}] \nn\\
  &\quad    -  \frac 14 \eta^{\mu\nu} 
  ([T^\a,T^\sigma][T_\a,T_\sigma]
  + 2[T^\a,T^{\ib}][T_\a,T_{\ib}]
   + [T^{\ib},T^{\jb}][T_{\ib},T_{\jb}])
\end{align}
and verify its conservation \eqref{T-cons}.
Separating again the components $T^\mu=\alpha t^\mu$  and $T_{\ib}^+=f(\tau)\cK_{\ib}^+$, we need the terms
\begin{align}
    [T^\mu,T^\nu]=&i\{\alpha t^\mu,\alpha t^\nu\}=-i\left[-\frac{1}{r^2R^2}\alpha^2\theta^{\mu\nu}+\frac{\alpha\alpha^\prime}{R}(t^\mu x^\nu-x^\mu t^\nu)\right]\\
    [T^\mu,T^\nu][T^\rho,T_\nu]=&-\frac{A^2}{r^4R^4}\theta^{\mu\nu}\theta^\rho_{\ \nu}-\frac{(A^\prime)^2}{4R^2}(t^\mu x^\nu-x^\mu t^\nu)(t^\rho x_\nu-t_\nu x^\rho)\nn\\
    &+\frac{AA^\prime}{2r^2R^3}\theta^{\mu\nu}(t^\rho x_\nu -t_\nu x^\rho)+\frac{AA^\prime}{2r^2R^3}(t^\mu x^\nu-x^\mu t^\nu)\theta^\rho_{\ \nu}\nn\\
    =&-\frac{A^2}{r^2R^2}\eta^{\mu\rho}-\frac{1}{R^2}\left(-{A^2}-\frac{{(A^\prime)^2(x_4^2+R^2)}}{4}-{AA^\prime}x_4\right)t^\mu t^\rho\nn\\
    &-\frac{1}{r^2R^4}\left({A^2}+\frac{(A^\prime)^2}{4}(R^2+x_4^2)+{AA^\prime}x_4\right)x^\mu x^\rho\nn\\
    \approx& -\frac{A^2}{r^2R^2}\eta^{\mu\rho}+\frac{1}{4R^2}\left(2A+\dot{A}\right)^2\left(t^\mu t^\rho-\frac{x^\mu x^\rho}{r^2R^2}\right)
\end{align}
denoting again $A=\alpha^2$. 
Note that this is symmetric in $\mu $ and $\rho$. We also need the $\eta^{AB}\mathcal{L}$ term, and the terms involving $\cK$, $(\mu,{\ib})$ and $({\ib},{\jb})$. We first compute
\begin{align}
    &[T^\mu,T^\nu][T_\mu,T_\nu]\approx 
     \frac{1}{r^2R^2}\left(-4A^2+2\frac{(A+\dot{A}/2)^2x_4^2}{r^2}\right)\ .
\end{align}
Furthermore, consider
\begin{align}
[T^\mu,T^{\ib}][T^\nu,T^{\ib}]&=-\alpha^2|f^\prime|^2C^2R^{-2}x^\mu x^\nu\\
[T^{\ib},T^{\jb}][T_{\ib},T_{\jb}]&=-|f|^4\Theta^{{\ib}{\jb}}\Theta_{{\ib}{\jb}}
\end{align}
with $\Theta$ being the antisymmetric NC parameter of $\cK$, i.e. $[\cK^{\ib},\cK^{\jb}]=i\Theta^{{\ib}{\jb}}$.
Then 
\begin{align}
     \mathcal{T}^{\mu\nu}
=&-\eta^{\mu\nu}\left(\frac{P}{2R^2}x_4^2+\frac{Q}{2R^2}-S\right)+r^2Pt^\mu t^\nu-\left(P+Qx_4^{-2}\right)\frac{x^\mu x^\nu}{R^2}
\end{align}
at late times, with
\begin{align}
    P&=\frac{(A+\dot A/2)^2}{r^2R^2}\\
    Q&={A|\dot f|^2C^2}\\
    S&=\frac{|f|^4\Theta^{\ib\jb}\Theta_{\ib\jb}}{4}=\frac{|f|^4\Lambda C^2}{4}
\end{align}
using \eqref{relation-K-const}.
\subsection*{Energy-momentum conservation}
We now verify e-m conservation
\begin{equation}\label{matrix-em-conserv}
    [T_A,\cT^{A\nu}]=[T_\mu,\cT^{\mu\nu}]+[T_{\ib},\cT^{\ib \nu}]=0
\end{equation}
with $T_\mu=\alpha t_\mu$. The first term is 
\begin{align}
    &i\{\alpha t_\mu,-\eta^{\mu\nu}\left(\frac{P}{2R^2}x_4^2+\frac{Q}{2R^2}-S\right)\}
    =i\alpha\frac{x_4x^\nu}{R^3} \left(\frac{\dot{P}+2 P}{2}+\frac{\dot Q}{2}x_4^{-2}-\dot SR^{2}x_4^{-2}\right)\nn\\
    &i\{\alpha t_\mu,r^2Pt^\mu t^\nu\}=-i\frac{\theta^{\mu\nu}t_\mu}{R^2}\alpha P+i\frac{1}{R^3}\alpha^\prime Px_4^2x^\nu\nn\\
    &=i\frac{x_4 P}{R^3}\left(\alpha+\dot\alpha\right)x^\nu 
    -i\{\alpha t_\mu,(P+Qx_4^{-2})x^\mu x^\nu/R^2\}\nn\\
    & =i\frac{x_4x^\nu}{R^3}\left[\alpha\left(-3P- \frac{\dot P}{2}-x_4^{-2}\left(\frac{\dot Q}{2}+3Q+\dot SR^2\right)\right)-\dot\alpha Qx_4^{-2}\right]\ .
\end{align}
The internal term can be decomposed as
\begin{equation}
    [T_{\ib},\mathcal{T}^{\ib \nu}]= [T_{\ib},[T^{\ib},T^\mu][T^\nu,T_{\mu}]]+[T_{\ib},[T^{\ib},T^{\jb}][T^\nu,T_{\jb}]]
\end{equation}
with first contribution
\begin{align}
    &[T_{\ib},[T^{\ib},\alpha t^\mu][\alpha t^\nu,\alpha t_\mu]]\nn\\
    &= -\frac{1}{2R}[fK_{\ib}^+,\alpha\bar{f}^\prime x_\mu \left(\frac{A}{r^2R^2}\theta^{\nu\mu}+\frac{A^\prime}{2R}(x^\nu t^\mu-t^\nu x^\mu)K^{\ib,-}\right) ]+h.c.\nn\\
    &= -\frac{1}{2R^2}[f \cK_{\ib}^+,\alpha\bar{f}^\prime \left(A+\dot{A}/2\right)x_4t^\nu \cK^{\ib,-}]+h.c.\nn\\ 
    &= -i\frac{C^2}{R^3}\alpha\left(A+\dot A/2\right)|\dot{f}|^2x_4^{-1}x^\nu-\frac{\alpha(A+\dot{A}/2)x_4}{2R^2}f\bar{f}^\prime[\cK_{\ib}^+,\cK^{\ib -}]t^\nu+h.c.\nn\\
    &= -i\frac{Q}{R^3}(\alpha+\dot\alpha)x_4^{-1}x^\nu-\frac{\alpha(A+\dot{A}/2)}{2R^2}[\cK_{\ib}^+,\cK^{\ib -}](f\bar{f}^\prime-\bar{f}f^\prime)t^\nu
\end{align}
and second contributon
\begin{align}
    &[T_{\ib},[T^{\ib},T^{\jb}][\alpha t^\nu,T_{\jb}]]=[T_{\ib},[T^{\ib},T^{\jb}]][\alpha t^\nu,T_{\jb}]+[T^{\ib},T^{\jb}][T_{\ib},[\alpha t^\nu,T_{\jb}]]\nn\\
    &= -i\alpha\frac{1}{2R}\Lambda C^2|f|^2f\bar{f}^\prime x^\nu +h.c.+\alpha\frac{|f|^2}{4R}\bar{f}^\prime\theta^{{\ib}{\jb},(+-)}[f\cK_{\ib}^+,\cK_{\jb}^-x^\nu]+h.c.\nn\\
    &= -i\frac{\alpha}{R}\dot{S}x_4^{-2}x^\nu+i\frac{\alpha}{R}\dot{S}x_4^{-2}x^\nu=0 \ .
\end{align}
Hence the matrix energy-momentum conservation equation \eqref{matrix-em-conserv} gives
\begin{align}
   0=&-\left[\alpha\left(-3P- \frac{\dot P}{2}-x_4^{-2}\left(\frac{\dot Q}{2}+3Q+\dot SR^2\right)\right)-\dot\alpha Qx_4^{-2}\right] -2\alpha R^2\dot{S}x_4^{-2}-Q(\alpha+\dot\alpha)x_4^{-2}\nn\\
   \overset{\frac{1}{\alpha}}{=}& \ 3P+ \frac{\dot P}{2}+x_4^{-2}\left(\frac{\dot{Q}}{2}+2Q+\dot{S}R^2\right) ,
\end{align}
and
\begin{equation}
    [\cK_{\ib}^+,\cK^{\ib -}](f\bar{f}^\prime-\bar{f}f^\prime) =0
\end{equation}
as in \eqref{K-condition-eom}.


\subsection*{Covariant conservation law}

The conservation
 law \eqref{current-cons-connection} can be written in terms of the Weitzenb\"ock connection $\nabla$  as 
\begin{align}
\label{div-weitze-levi}
    \nabla^\mu V_\mu = \gamma^{\mu\nu}\nabla_\nu V_\mu 
     &= \gamma^{\mu\nu}(\partial_\nu V_\mu - \tensor{\Gamma}{_\nu_\mu^\rho} V_{\rho}) \nn\\
     &= \gamma^{\mu\nu}\partial_\nu V_\mu 
     + \frac{\rho^2}{\sqrt{|G|}} \partial_\sigma(\sqrt{|G|} G^{\rho\sigma}) V_{\rho} \nn\\
     &= \frac{\rho^2}{\sqrt{|G|}} \partial_\sigma(\sqrt{|G|} G^{\rho\sigma} V_{\rho}) \nn\\
     &= \rho^2\nabla^{(G) \mu} V_{\mu}
\end{align}
using the identity (9.2.47) in \cite{Steinacker:2024unq}
\begin{align}
\gamma^{\mu\nu}\tensor{\Gamma}{_\nu_\mu^\rho} = -\frac{\rho^2}{\sqrt{|G|}} \partial_\sigma(\sqrt{|G|} G^{\rho\sigma})
\end{align}
for the Weizenb\"ock connection.

\section{Frame contribution of extra dimensions}

In the presence of time-dependent 
(e.g. internally rotating)
extra dimensions 
$T^{\ib+ } =  f(\tau) \cK^{\ib+}$, 
the kinetic term acquires an extra term, which modifies the effective geometry. This can be taken into account by an extra (Euclidean) frame contribution of the form 
\begin{align}
    E^{{\ib}+,\mu} &= \{f(x_4) \cK^{\ib+},x^\mu\} 
     \approx r \dot{f} \cK^{\ib+} u^\mu \ .
\end{align}
Note that this is not a classical function but $\hs$ valued; moreover, we could generically not find local normal coordinates adapted to more than 4 frame fields.
Therefore this term should be negligible
compared with the frame on $\cM$, which  requires that 
\begin{align}
   r\dot f|\cK| & \ll \a \sinh\tau  \sim e^{(\varepsilon+1)\tau}\ .
\end{align}
This holds at late times provided $f \ll e^{(\varepsilon+1)\tau}$, which is indeed satisfied on the classical solutions and on the preferred physical solution $\varepsilon=-\frac{3}{4}$.

\section{Relation between the $\cK$ constants}

We observe that if the equations
\begin{equation}
    [\cK_{\ib},[\cK^{\ib},\cK_{\jb}]]=\Lambda\cK_{\jb},\qquad\cK_{\ib}\cK^{\ib}=C^2\one
\end{equation}
are satisfied, then the following relation holds
\begin{equation}\label{relation-K-const}
     \Theta^{{\ib}{\jb}}\Theta_{{\ib}{\jb}}=\Lambda C^2
\end{equation}
where $\Theta^{\ib \jb}=-i[\cK^{\ib},\cK^{\jb}]$.
This can be seen by an explicit computation:
\begin{align}
    \Lambda C^2&=\Lambda \cK_{\jb}\cK^{\jb}=\cK_{\jb}[\cK_{\ib},[\cK^{\ib},\cK^{\jb}]]=2C^4-2\cK_{\jb }\cK_{\ib}\cK^{\jb}\cK^{\ib}\nn\\
    &=-(-\cK_{\jb}\cK_{\ib}\cK^{\ib}\cK^{\jb}-\cK_{\ib}\cK_{\jb}\cK^{\jb}\cK^{\ib}+\cK_{\ib}\cK_{\jb}\cK^{\ib}\cK^{\jb}+\cK_{\jb}\cK_{\ib}\cK^{\jb}\cK^{\ib})\nn\\
    &=\Theta_{{\ib}{\jb}}\Theta^{{\ib}{\jb}}\ .
\end{align}

\bibliography{twistor}
\bibliographystyle{JHEP-2}
\end{document}